\documentclass[twocolumn]{aastex63}
\usepackage[utf8]{inputenc}
\usepackage{graphics,graphicx}
\usepackage{epsfig}
\usepackage{url}
\usepackage{wrapfig}
\usepackage{float}
\usepackage{subfigure}
\usepackage{xcolor}

\shortauthors{J.-T. Li et al.}
\shorttitle{HIERACHY II: Project Design}

\begin{document}

\title{Probing the \ion{He}{2} re-Ionization ERa via Absorbing \ion{C}{4} Historical Yield (HIERACHY) II: Project Design, Current Status, and Examples of Initial Data Products}

\author[0000-0001-6239-3821]{Jiang-Tao Li}
\affiliation{Purple Mountain Observatory, Chinese Academy of Sciences, 10 Yuanhua Road, Nanjing 210023, People’s Republic of China}

\author{Xiaodi Yu}
\affiliation{Department of Astronomy, Tsinghua University, Beijing 100084, People’s Republic of China}
\affiliation{Purple Mountain Observatory, Chinese Academy of Sciences, 10 Yuanhua Road, Nanjing 210023, People’s Republic of China}

\author[0009-0006-7138-2095]{Huiyang Mao}
\affiliation{Purple Mountain Observatory, Chinese Academy of Sciences, 10 Yuanhua Road, Nanjing 210023, People’s Republic of China}

\author[0000-0001-8892-8759]{Hanxiao Chen}
{\affiliation{School of Astronomy and Space Science, Nanjing University, Nanjing, Jiangsu 210093, People’s Republic of China}
\affiliation{Key Laboratory of Modern Astronomy and Astrophysics, Nanjing University, Ministry of Education, Nanjing 210093, People’s Republic of China}}

\author{Tiancheng Yang}
\affiliation{School of Astronomy and Space Science, Nanjing University, Nanjing, Jiangsu 210093, People’s Republic of China}
\affiliation{Key Laboratory of Modern Astronomy and Astrophysics, Nanjing University, Ministry of Education, Nanjing 210093, People’s Republic of China}

\author[0000-0002-2941-646X]{Zhijie Qu}
\affiliation{Department of Astronomy \& Astrophysics, The University of Chicago, 5640 S. Ellis Ave., Chicago, IL 60637, USA}

\author{Fuyan Bian}
\affiliation{European Southern Observatory, Alonso de C\'{o}rdova 3107, Casilla 19001, Vitacura, Santiago 19, Chile}

\author[0000-0001-6239-3821]{Joel N. Bregman}
\affiliation{Department of Astronomy, University of Michigan, 311 West Hall, 1085 S. University Ave, Ann Arbor, MI, 48109-1107, USA}

\author{Zheng Cai}
\affiliation{Department of Astronomy, Tsinghua University, Beijing 100084, People’s Republic of China}

\author[0000-0003-3310-0131]{Xiaohui Fan}
\affiliation{Steward Observatory, University of Arizona, 933 North Cherry Avenue, Tucson, AZ 85721, USA}

\author[0000-0002-2853-3808]{Taotao Fang}
\affiliation{Department of Astronomy and Institute of Theoretical Physics and Astrophysics, Xiamen University, Xiamen, Fujian 361005, People’s Republic of China}

\author{Li Ji}
\affiliation{Purple Mountain Observatory, Chinese Academy of Sciences, 10 Yuanhua Road, Nanjing 210023, People’s Republic of China}

\author{Zhiyuan Ji}
\affiliation{Steward Observatory, University of Arizona, 933 North Cherry Avenue, Tucson, AZ 85721, USA}

\author[0000-0001-9487-8583]{Sean D. Johnson}
\affiliation{Department of Astronomy, University of Michigan, 311 West Hall, 1085 S. University Ave, Ann Arbor, MI, 48109-1107, USA}

\author{Guoliang Li}
\affiliation{Purple Mountain Observatory, Chinese Academy of Sciences, 10 Yuanhua Road, Nanjing 210023, People’s Republic of China}

\author[0000-0003-3762-7344]{Weizhe Liu}
\affiliation{Steward Observatory, University of Arizona, 933 North Cherry Avenue, Tucson, AZ 85721, USA}

\author[0000-0002-6270-8851]{Ying-Yi Song}
\affiliation{David A. Dunlap Department of Astronomy \& Astrophysics, University of Toronto, 50 St. George Street, Toronto, ON M5S 3H4, Canada}
\affiliation{Dunlap Institute for Astronomy \& Astrophysics, University of Toronto, 50 St. George Street, Toronto, ON M5S 3H4, Canada}

\author[0000-0002-7633-431X]{Feige Wang}
\affiliation{Steward Observatory, University of Arizona, 933 North Cherry Avenue, Tucson, AZ 85721, USA}

\author{Tao Wang}
\affiliation{School of Astronomy and Space Science, Nanjing University, Nanjing, Jiangsu 210093, People’s Republic of China}
\affiliation{Key Laboratory of Modern Astronomy and Astrophysics, Nanjing University, Ministry of Education, Nanjing 210093, People’s Republic of China}

\author[0000-0002-9373-3865]{Xin Wang}
\affil{School of Astronomy and Space Science, University of Chinese Academy of Sciences (UCAS), Beijing 100049, People’s Republic of China}
\affil{National Astronomical Observatories, Chinese Academy of Sciences, Beijing 100101, People’s Republic of China}
\affil{Institute for Frontiers in Astronomy and Astrophysics, Beijing Normal University,  Beijing 102206, People’s Republic of China}

\author{Christina Williams}
\affiliation{NSF’s National Optical-Infrared Astronomy Research Laboratory, 950 North Cherry Avenue, Tucson, AZ 85719, USA}

\author{Mingxuan Xu}
\affiliation{Purple Mountain Observatory, Chinese Academy of Sciences, 10 Yuanhua Road, Nanjing 210023, People’s Republic of China}

\author[0000-0001-5287-4242]{Jinyi Yang}
\affiliation{Steward Observatory, University of Arizona, 933 North Cherry Avenue, Tucson, AZ 85721, USA}

\author[0000-0001-7254-219X]{Yang Yang}
\affiliation{Purple Mountain Observatory, Chinese Academy of Sciences, 10 Yuanhua Road, Nanjing 210023, People’s Republic of China}

\author{Xianzhong Zheng}
\affiliation{Purple Mountain Observatory, Chinese Academy of Sciences, 10 Yuanhua Road, Nanjing 210023, People’s Republic of China}

\correspondingauthor{Jiang-Tao Li}
\email{pandataotao@gmail.com}

\begin{abstract}
The \ion{He}{2} reionization epoch is expected to take place at $z\sim3-5$. In this stage, the helium and metals in the inter-galactic medium (IGM) are further ionized with additional contributions from harder non-stellar sources, and some large-scale gravitationally bound systems approach virialization. The ``Probing the \ion{\textbf{H}e}{2} re-\textbf{I}onization \textbf{ER}a via \textbf{A}bsorbing \ion{\textbf{C}}{4} \textbf{H}istorical \textbf{Y}ield (HIERACHY)'' program utilizes high- and medium-resolution spectra of bright background quasars at $z\approx3.9-5.2$ to investigate Ly$\alpha$, \ion{C}{4}, and other metal absorption lines during this epoch. Additionally, we employ narrow-band imaging to search for Ly$\alpha$ emitters associated with \ion{C}{4} absorbers, alongside multi-wavelength observations to identify and study particularly intriguing cases. In this paper, we present the design of the HIERACHY program, its current status, major scientific goals, and examples of initial data products from completed Magellan/MIKE, MagE spectroscopy, and MDM imaging observations. We also provide a brief outlook on future multi-wavelength observations that may significantly impact the related science.
\end{abstract}


\section{Introduction} \label{sec:Background}

In cosmic reionization theory, star-forming galaxies play a crucial role in the reionization of \ion{H}{1} and \ion{He}{1} at $z\gtrsim6$, converting most of the helium in the intergalactic medium (IGM) to \ion{He}{2} (e.g., \citealt{Fan06}). However, the UV spectra of young stars are typically not hard enough to doubly ionize helium. Instead, quasars are considered the most likely ionizing sources responsible for \ion{He}{2} reionization, which likely began at $z\gtrsim5$ and was completed by $z\gtrsim3$ (e.g., \citealt{Davidsen96,Oh01,McQuinn09,FaucherGiguere09,McQuinn16}). During this \ion{He}{2} Epoch of Reionization (EoR), the heat input from \ion{He}{2} ionization could dominate the thermal balance of the IGM (e.g., \citealt{Becker11,Madau15}). The gas recycling within galactic ecosystems is strongly influenced by the \ion{He}{2} reionization process. Many aspects of this process, such as the relative importance of quasars, the UV background, and other components in providing ionizing photons, as well as the redshift evolution of temperature, comoving mass density, and column density distribution of different ions in the IGM, remain poorly constrained observationally (e.g., \citealt{Schaye00,Becker11}).

The cosmic reionization history is also closely related to the formation, growth, and feedback of supermassive black holes (SMBHs) in the center of galaxies. SMBHs as massive as $M_{\rm SMBH}\sim10^9\rm~M_\odot$ appeared as early as $z>7.5$ (or only $\sim700$ million years after the Big Bang, e.g., \citealt{Banados18,Yang20,Wang21}), and less massive ones were discovered at even higher redshifts with the JWST (e.g., \citealt{Juodzbalis23,Larson23,Maiolino24}). Some of these SMBHs grew to a mass as large as the most massive ones in the local universe before the end of the Hydrogen EoR ($z>6$, $M_{\rm SMBH}\sim10^{10}\rm~M_\odot$, \citealt{Wu15}), and many may keep accreting actively at lower redshifts, contributing significantly to the \ion{He}{2} reionization (e.g., \citealt{McQuinn09,FaucherGiguere09}).

Cosmological structure formation and evolution are controlled by processes happening on a wide range of spatial scales, from stellar sources (e.g., via stellar wind or supernovae, SNe) and SMBHs living at the center of galaxies (e.g., via AGN feedback; both launched at $\lesssim10^{-6}\rm~pc$) to massive galaxy clusters (e.g., via merger shock; $\sim10^6\rm~pc$) or even larger scale structures (e.g., cosmic filaments). While proto-clusters are confirmed to exist as early as in the Hydrogen EoR (e.g., \citealt{Toshikawa14,Higuchi19,Hu21,Wang24a}), the earliest mature galaxy clusters with virialized hot intra-cluster medium (ICM) detectable in X-ray and Sunyaev–Zel'dovich (SZ) signals are only confirmed after the ending of the \ion{He}{2} EoR ($z\lesssim3$, e.g., \citealt{WangT16,Tozzi22}). How the formation of this large-scale gravitationally bound structure was related to SMBH growth and feedback, as well as \ion{He}{2} reionization, is critical to our understanding of the co-evolution of SMBHs, galaxies, and their large-scale environments.

The \ion{C}{4} doublet at rest frame $\lambda\lambda$1548.2, 1550.8$\text{\AA}$ from intergalactic absorbers in the spectra of bright background quasars at redshift $z\gtrsim4$ is probably the best tracer of the \ion{He}{2} reionization (e.g., \citealt{Yu21}). This is because: (1) The ionization potential of \ion{C}{4} (64.5~eV) is comparable to \ion{He}{2} (54.4~eV) \citep{Kramida20}. Therefore \ion{C}{4} lines trace the same gas phase dominated by \ion{He}{2}, which could be at least partially hot gas instead of only photo-ionized cool gas (e.g., \citealt{Pettini82}). (2) The \ion{C}{4} doublets could be easily detected, they are among the strongest absorption lines in the rest frame UV band (e.g., \citealt{Young82,Hasan20}). (3) The doublets are easy to identify, as their wavelength and oscillator strength are well-determined in atomic physics and there are no other strong lines in a similar wavelength range (e.g., \citealt{Feibelman83,Petitjean04}). (4) The \ion{C}{4} doublet transitions exhibit significantly longer wavelengths than the Hydrogen Ly$\alpha$ line ($\lambda$1215.67$\text{\AA}$), thus they are not affected by the Ly$\alpha$ forest when the redshift of the absorber is not much lower than the quasar (e.g., \citealt{Pettini03}). The lower limit of the usable redshift range of the absorber can be described as: $z_{\rm abs}>\frac{\lambda_{\rm Ly\alpha}}{\lambda_{\rm C IV}}z_{\rm AGN}+\frac{\lambda_{\rm Ly\alpha}}{\lambda_{\rm C IV}}-1$, where $z_{\rm abs}$ and $z_{\rm AGN}$ are the redshifts of the absorber and the background AGN, while $\lambda_{\rm Ly\alpha}$ and $\lambda_{\rm C IV}$ are the wavelength of Ly$\alpha$ and the \ion{C}{4} doublet. Considering the uncertainties, for AGNs at $z_{\rm AGN}\sim4$, we can in principle detect \ion{C}{4} absorbers at $z_{\rm abs}\gtrsim3$. 

There are many \ion{C}{4} absorption line studies using spectra of quasars in a broad redshift range (e.g., \citealt{Songaila01,Pettini03,Simcoe11,Cooksey13,DOdorico13,DOdorico22,Hasan20,Davies23}), but most of them have a small sample size, low signal-to-noise ratio ($\rm S/N$), or poor spectral resolution. With a typical equivalent width (EW) detection limit of $>30\rm~m\text{\AA}$ in most of the previous observations, we cannot detect weak absorbers with $\log N{\rm(C{\small~IV})/cm^2}\lesssim13$ (e.g., \citealt{Puech18}; see \citealt{Yu24} for detailed comparisons). Furthermore, the blending of different absorption components may be fatal in the measurement, identification, or even detection of the absorption lines. Stronger absorbers may also be highly contaminated by outflows instead of the IGM (e.g., \citealt{Nestor08,Wang18}). We therefore need high $\rm S/N$, high resolution optical spectra of more bright quasars at $z\gtrsim4$ to systematically study the \ion{He}{2} reionization process.

In this paper, we introduce a new program, which studies the \ion{He}{2} reionization via \ion{C}{4} and other metal or Ly$\alpha$ absorption lines and other related processes in the same redshift range. We introduce the design of the program in \S\ref{sec:ProjectDesign}, including the quasar sample (\S\ref{subsec:Sample}), the completed spectroscopy  (\S\ref{subsec:MagellanMIKE}, \S\ref{subsec:MagellanMagE}) observations, and initial comparison to other samples (\S\ref{subsec:CompareOthers}). We introduce the data reduction procedures of our high-resolution spectroscopy observations in \S\ref{sec:MIKEDateReduction}. We then discuss the major scientific goals of the program and present some examples of initial data products in \S\ref{sec:SciGoals}. A summary of the current status of the program and a prospect on future multi-wavelength observations will be presented in \S\ref{sec:Summary}. Throughout the paper, we adopt a cosmological model with $H_{\rm 0}=70\rm~km~s^{-1}~Mpc^{-1}$, $\Omega_{\rm M}=0.3$, $\Omega_{\rm \Lambda}=0.7$, and $q_{\rm 0}=-0.55$. All errors are quoted at a confidence level of $1~\sigma$ unless specifically noted.

\section{Project Design} \label{sec:ProjectDesign}

We herein introduce the ``Probing the \ion{\textbf{H}e}{2} re-\textbf{I}onization \textbf{ER}a via \textbf{A}bsorbing \ion{\textbf{C}}{4} \textbf{H}istorical \textbf{Y}ield (HIERACHY)'' program. The primary scientific goal of the HIERACHY program is to study the \ion{He}{2} reionization and other relevant processes at redshift $z\sim3-5$, such as AGN outflow and the formation of large-scale gravitationally bound systems, i.e., (proto-)clusters. We are aiming at answering a few key questions related to the \ion{He}{2} reionization processes, including but not limited to: (1) What are the properties (ionization state, temperature, metallicity, etc.) of the IGM in the \ion{He}{2} EoR? (2) What are the major ionizing sources of the IGM in the \ion{He}{2} EoR? (3) How do AGN, galaxies, and the multi-phase CGM co-evolve with each other? How do these processes affect and get affected by the heating and reionization of the IGM? (4) When and how do large-scale gravitationally bound systems, such as (proto-)clusters, form and get virialized?

The HIERACHY program is comprised of a few independent but closely related components: (1) Identify intervening \ion{C}{4} absorbers (\S\ref{subsec:SciGoalzEvolvNCIVFunction}) and other associated metal (\S\ref{subsec:SciGoalMultiphaseIGM}) or Ly$\alpha$ absorbers (\S\ref{subsec:SciGoalLyaForestIGMEoS}), based on optical spectra of the brightest background quasars typically at $z\gtrsim4$. (2) Search for Ly$\alpha$ emitters (LAE) associated with the identified \ion{C}{4} absorbers, based on optical narrow-band images (\S\ref{subsubsec:SciGoalCIVHostGalaxy}). Here the \ion{C}{4} absorbers could be from either step~(1) or the archive (e.g., SDSS). Another by-product of this step is the search for extended emission line nebulae associated with strong \ion{C}{4} absorbers. (3) Use follow-up multi-object spectroscopy observations either in optical or radio bands to study properties of the member galaxies of the identified (proto-)cluster candidates (\S\ref{subsubsec:SciGoalProtoCluster}). (4) Search for the X-ray or SZ signal from the ICM of massive clusters at $z\gtrsim3$ (\S\ref{subsubsec:SciGoalProtoCluster}). 

A pilot case study of the HIERACHY program, as a test of the Magellan/MIKE observations (\S\ref{subsec:MagellanMIKE}), has already been published in \citet{Yu21} (HIERACHY~I), which present our discovery of a fast ($v\sim-6500\rm~km~s^{-1}$) outflow in a $z\approx4.7$ quasar. In the present paper, we will introduce the status and some initial data products in step (1), while more quasar absorption line observations, as well as other follow-up observations in steps (2-4), are still ongoing. In a companion paper (\citealt{Yu24}; HIERACHY~III), we will publish our \ion{C}{4} absorber catalogue obtained solely from the Magellan/MIKE observations.

\subsection{Quasar Samples} \label{subsec:Sample}

We use two types of quasar samples in the HIERACHY program: (1) the brightest background quasars which are used for both the high to medium resolution optical spectroscopy observations taken with the Magellan telescopes (\S\ref{subsec:MagellanMIKE}, \S\ref{subsec:MagellanMagE}) and many multi-wavelength follow-up observations; and (2) quasars from the archive (mainly SDSS) which are typically at lower redshifts and/or have lower resolution spectra. The latter sample is only used to identify the strongest \ion{C}{4} absorbers, which could be candidates for multi-wavelength follow-up observations. It is not used for statistical analysis studying the \ion{He}{2} reionization.

We need the brightest background quasars during or before the \ion{He}{2} reionization epoch to obtain high-resolution optical (rest frame UV) spectroscopy observations to reach a low enough detection limit of the absorption lines in order to study the IGM/CGM (\S\ref{subsec:MagellanMIKE}, \ref{subsec:MagellanMagE}). Our initial sample is based on the collection of known $z\geq4.5$ quasars from \citet{WangF16}, plus some newly discovered quasars from different later literatures (e.g., \citealt{Wang18,Yang19,Wolf20}). The original sample includes $>10^3$ $z\geq4.5$ quasars with spectroscopic redshift and multi-band photometry. Its multi-wavelength (e.g., X-ray and radio) properties are further studied in \citet{Li21a,Li21b}. We typically select the brightest quasars in this sample visible from the Magellan telescopes in the southern sky (observed with two instruments, as will be described in details in \S\ref{subsec:MagellanMIKE} and \ref{subsec:MagellanMagE}). Since the program is carried out over multiple Magellan observational seasons, and in each allocated night, the objects are chosen based on both their properties and visibility, our sample is far from complete. We describe details of the selection criteria based on our $z\geq4.5$ quasar sample and the completed observations in \S\ref{subsec:MagellanMIKE} and \ref{subsec:MagellanMagE}. In addition to this $z\geq4.5$ sample, we also add a few bright quasars at slightly lower redshifts, which could cover \ion{C}{4} absorbers down to $z\sim3$. The sample is being updated continuously based on latest quasar surveys and our completed observations.

In addition to the relatively high resolution optical spectra taken with the Magellan telescopes, we also select a sample from the SDSS DR7 catalogue of high-$z$ \ion{C}{4} absorbers \citep{Cooksey13}. The SDSS spectra used in \citet{Cooksey13} have a resolution of $R\approx2,000$, which is only useful to identify the strongest \ion{C}{4} absorbers (e.g., Fig.~\ref{fig:J1509MDMimages}b). \citet{Cooksey13} identified $>16,000$ \ion{C}{4} absorption systems with a limiting equivalent width of $W_{r}\approx0.6\rm~\text{\AA}$ in the redshift range of $z=1.46-4.55$. We use a few different criteria to select some strong and complex \ion{C}{4} absorption systems as tracers of candidates of foreground galaxy overdensities or (proto-)clusters. These candidates will be used for follow-up narrow-band imaging observations to identify LAEs associated with the \ion{C}{4} absorbers. An example of the narrow-band imaging observations anchored to a strong \ion{C}{4} absorber detected with the SDSS spectra is presented in \S\ref{subsubsec:SciGoalCIVHostGalaxy}, while more details about the sample selection criteria and our narrow-band imaging observations will be presented in follow-up papers.

\subsection{Magellan/MIKE High-Resolution Spectroscopy Observations} \label{subsec:MagellanMIKE}

In the first step of the HIERACHY program, we take high-resolution optical spectra (cover the rest frame UV band) of the brightest high-$z$ quasars with the Magellan Inamori Kyocera Echelle (MIKE) spectrograph on the 6.5m Magellan II Clay telescope \citep{Bernstein03}. MIKE is the double arm high-resolution optical Echelle spectrograph installed on the Nasmyth East port of the Clay telescope. It has two arms (blue and red) which provide full wavelength coverage of about $3350-5000\rm~\text{\AA}$ (blue) and $4900-9500\rm~\text{\AA}$ (red) in its standard configuration. The peak efficiency of the two arms are $\sim19\%$ (blue) and $\sim14\%$ (red), respectively. In our sample, since the Ly$\alpha$ emission line of the quasars has been redshifted to $>5950\rm~\text{\AA}$, we will mostly use the data taken with the red arm to identify the \ion{C}{4} absorbers and most of the metal line absorbers, while the blue arm will be used to study the Ly$\alpha$ forest at $z\lesssim3$, as well as some blue metal lines. MIKE has a few single slits or pair of slits with a width ranging from $0.35^{\prime\prime}$ to $2^{\prime\prime}$. The peak spectral resolution with the narrowest $0.35^{\prime\prime}$ slit appear at $\sim7000\rm~\text{\AA}$ ($R\sim65,000$). In most of the cases, with limited seeing conditions (typically $\sim0.6^{\prime\prime}$), we choose a $0.7^{\prime\prime}\times5.0^{\prime\prime}$ entrance slit and a $2\times2$ binning of the detector (pixel scale $\approx0.1\rm~\text{\AA}~pixel^{-1}$ after binning; hereafter we refer to the binned pixel as pixel). This setting results in a spectral resolution of $R\sim32,000$ for the red channel and $R\sim41,000$ for the blue channel.

The real observation settings (the slit width, the exposure time, etc.) are slightly adjusted for different targets based on the weather (e.g., seeing), the location (airmass) of the objects, the moon phase, and the available observation time each night. For example, in one case only (J100114+211514 at $z\approx3.96$), we choose the $1^{\prime\prime}\times5.0^{\prime\prime}$ entrance slit, which resulted in a resolution of $R\sim23,000$ for the red channel. Most of the objects with a short exposure time and low $\rm S/N$ were observed at the end of the night and did not have sufficient observation time. In most of the observations, we adopt 2-3.5~hours of effective exposure time (typically separated into a few 1200s or 1800s exposures) for each object, which resulted in a $\rm S/N\sim20-40\rm~pixel^{-1}$ at the continuum redder than the Ly$\alpha$ emission line. 

From the 2019A to 2022A semester, we completed Magellan/MIKE observations of 26 quasars in the redshift range of $z\approx3.9-5.2$ (Table~\ref{table:MIKEsample}). The lower limit of the redshift ensures the coverage of the \ion{He}{2} EoR with the \ion{C}{4} absorption line ($z\gtrsim3$). This is also the redshift where the halos of proto-clusters are expected to transition from a regime where the gas inflow in a $M_{\rm halo}\gtrsim10^{12}\rm~M_\odot$ massive halo could still be cold to a regime where all such accreted gas is shock-heated to the virial temperature (e.g., \citealt{Overzier16}). This redshift range is thus critical for searching for the first mature clusters with virialized hot ICM. On the other hand, the upper limit on the redshift ($z\sim5.2$) ensures that any \ion{C}{4} absorbers in our major redshift range of interest will fall at longer wavelengths than the Ly$\alpha$ line of the quasar and be covered by MIKE. 

We need the brightest quasars as the background source to reach a low enough \ion{C}{4} column density detection limit, which requires both a high S/N and a high spectral resolution. Our original $z\geq4.5$ quasar catalogue does not contain an $i$-band magnitude which peaks at $\gtrsim7000\rm~\text{\AA}$ covering the wavelength range of interest for all the quasars \citep{WangF16,Li21b}. We then typically select the quasar sample based on their $J$-band magnitude (with an initial criterion of $\rm mag_J<17.5$ but also include those without a published $J$-band magnitude). We then search for the $i$-band magnitude of individual selected quasars on SIMBAD\footnote{https://simbad.u-strasbg.fr/simbad/}, and only select those with $\rm mag_i\lesssim19.0$ or those without a published $i$-band magnitude but with a $J$-band magnitude of $\rm mag_J\lesssim17.0$ or a Gaia Red Photometer (RP) magnitude \citep{Gaia16} of $\rm R_p\lesssim18.0$. Within a moderate amount of exposure time (e.g., $\sim3\rm~hours$), we can typically reach a $\rm S/N\gtrsim20\rm~pixel^{-1}$ at a full resolution of $R\gtrsim3\times10^4$, allowing us to detect \ion{C}{4} absorbers with a column density as low as $\log N{\rm(C{\small~IV})/cm^2}\sim12$ (see an example in Fig.~\ref{fig:ExampleMIKESpec}, as well as an initial case study in HIERACHY~I). For the 26 quasars in our Magellan/MIKE sample, 17 have a $\rm S/N_{25\%}\gtrsim20\rm~pixel^{-1}$, where $\rm S/N_{25\%}$ is the 25th percentile $\rm S/N$ in the $\rm S/N$ probability distribution functions (PDFs; Fig.~\ref{fig:ExampleMIKESpec}b). The $\rm S/N$ PDF is calculated using pixels only in the band used to search for the \ion{C}{4} absorbers (the purple part in Fig.~\ref{fig:ExampleMIKESpec}a). The minimum wavelength of this band is approximately at the Ly$\alpha$ emission line, while the maximum wavelength is set to $5000\rm~km~s^{-1}$ blueward of the \ion{C}{4}~$\lambda1550\rm~\text{\AA}$ emission line of the quasar. The detailed setting of this \ion{C}{4} absorber searching band of each quasar is presented in HIERACHY~III. $\rm S/N_{25\%}$ is a characteristic lower limit on the $\rm S/N$ of all the usable pixels in the redshift range of interest. Similarly, we also define $\rm S/N_{75\%}$ and use these two characteristic $\rm S/N$ to describe the quality of the data (listed in Table~\ref{table:MIKEsample}). We present an example MIKE spectrum in Fig.~\ref{fig:ExampleMIKESpec}, in order to show the definition of $\rm S/N_{25\%}$ and the \ion{C}{4} absorber searching band.

\begin{figure*}[!h]
\begin{center}
\includegraphics[width=0.75\textwidth]{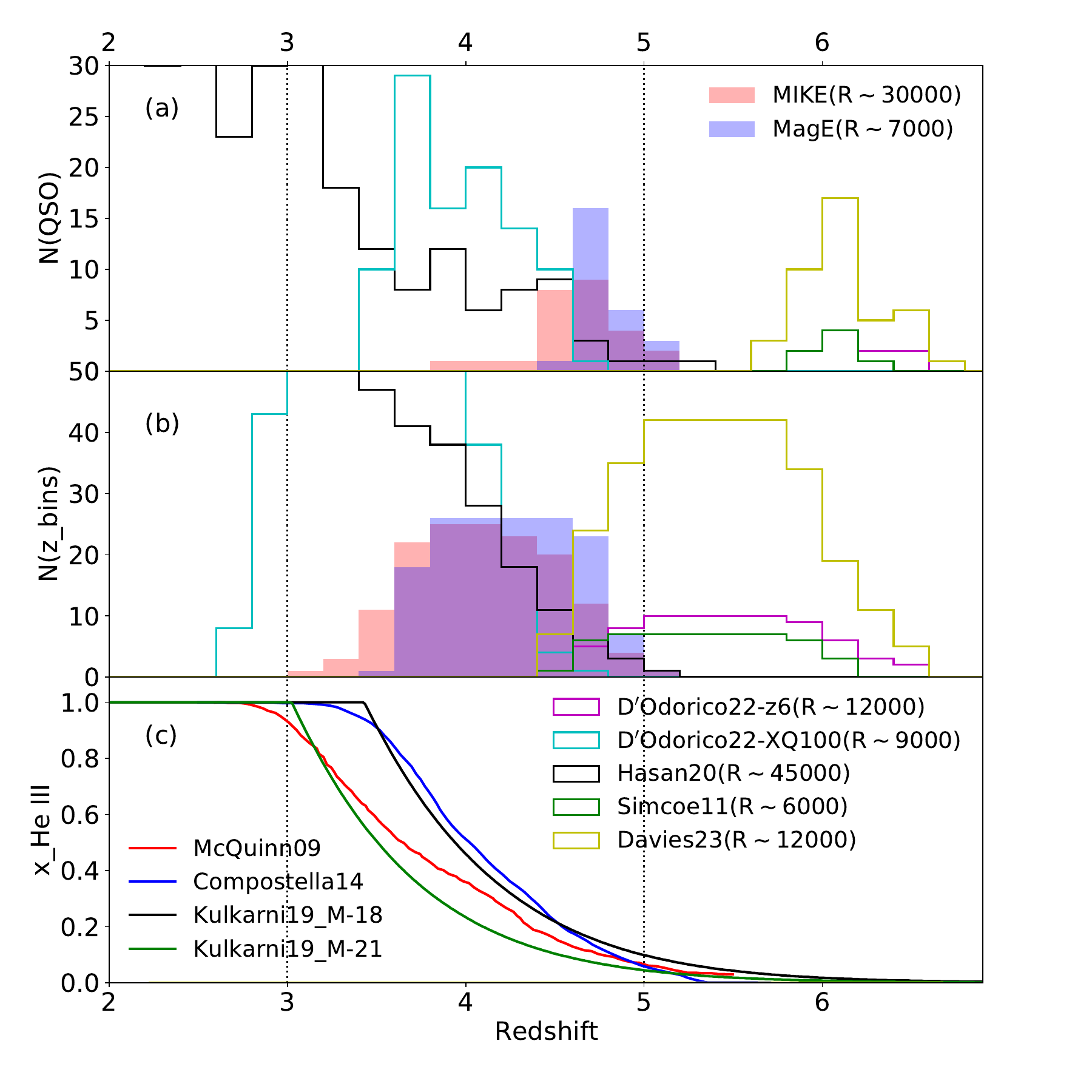}
\caption{Comparison of the Magellan/MIKE and MagE samples from the HIERACHY program to other similar samples and theoretical models. The other samples included for comparisons are \citet{Hasan20}'s Keck/HIRES and VLT/UVES spectra of 369 quasars at $z\sim1.1-5.3$, with a typical spectral resolution of $R\sim45,000$ (Hasan20); \citet{DOdorico22}'s VLT/X-shooter observations of 10 quasars at $z\sim5.8-6.5$ (D'Odorico22-z6) and 100 quasars at $z\sim3.5-4.8$ (D'Odorico22-XQ100), with a typical spectral resolution of $R\sim9,000$; \citet{Simcoe11}'s Magellan/FIRE observations of seven quasars at $z\sim5.8-6.3$, with a typical spectral resolution of $R\sim6,000$ (Simcoe11); \citet{Davies23}'s VLT/X-shooter spectra of 42 quasars at $z\sim5.8-6.6$ (30 with new observations and 12 from archive). Panel~(a) shows the number of quasars at the given redshift bins (bin size $\Delta z=0.2$) of each sample. Panel~(b) shows the total number of redshift bins which could be used to search for the \ion{C}{4} absorbers. The corresponding redshift range is typically from the Ly$\alpha$ emission line of the quasar to $\Delta v = 5000\rm~km~s^{-1}$ blueward of the \ion{C}{4} emission line of the quasar (e.g., Fig.~\ref{fig:ExampleMIKESpec}). Panel~(c) shows various theoretical predictions of the volume-averaged \ion{He}{3} fraction \citep{McQuinn09,Compostella14,Kulkarni19}. The two vertical dotted lines mark the approximate range of the expected epoch of \ion{He}{2} reionization which is best probed with the HIERACHY program.} \label{fig:sample}
\end{center}
\end{figure*}


\begin{table*}
\begin{center}
\caption{Magellan/MIKE sample of bright quasars at $z=3.9-5.2$} 
\begin{tabular}{lcccccccc}
\hline\hline
 Quasar & $z$ & $\rm mag_{i}$/$\rm mag_{J}$/$\rm R_{p}$ & $t_{\rm exp}$ & $\rm S/N~pixel^{-1}$ & Date & $P_{\rm moon}$ & Mode & $\Delta z_{\rm C{\scriptscriptstyle~IV}}$ \\
 & & & (hr) & $25\%/75\%$ & & (\%) & & \\ 
\hline 
J013127-032100	&	5.18	& 	18.4/-/-	 	&	3	&	22/28	&	Aug.16-17 2021 & 68 &	remote	& 3.8527-5.0614\\ 
J234433+165316	&	5.00	& 	18.6/-/-	 	&	1.2	&	8/11	&	Aug.17 2021	& 78 &	remote	& 3.6971-4.8820\\ 
J001115+144601	&	4.96	& 	18.3/-/-	 	&	2.2	&	23/27	&	Aug.16 2021	& 68 &	remote	& 3.6799-4.8606\\ 
J045427-050049	&	4.93	& 	18.6/-/-	 	&	0.8	&	12/14   &	Dec.28 2019	& 9 &	on site	& 3.6250-4.7917\\ 
J230429-313426	&	4.84	& 	-/-/17.7	 	&	3.0	&	29/35	&	Sep.25 2021	& 76 &	remote	& 3.5857-4.7426\\	
J111054-301129	&	4.83	& 	17.4/-/-	 	&	3.3	&	40/46	&	Mar.01 2022	& 0 &	remote	& 3.5615-4.7123\\ 
J094604+183539	&	4.80	& 	18.3/17.2/-	    &	1.3	&	6/8	    &	Dec.27 2019	& 4 &	on site	& 3.5544-4.7033\\ 
J030722-494548	&	4.78	& 	-/16.4/-	 	&	2.4	&	38/44	&	Dec.27 2019	& 4 &	on site	& 3.5647-4.7163\\ 
J091655-251145	&	4.77	& 	17.2/-/-	 	&	2.5	&	37/42	&	Mar.01 2022	& 0 &	remote	& 3.5686-4.7211\\ 
J145147-151220	&   4.76	& 	-/16.1/-	 	&	2	&	35/41	&	Apr.28 2019	& 29 &	on site	& 3.5229-4.6639\\ 
J014741-030247	&	4.75	& 	18.6/-/-	 	&	2.8	&	17/21	&	Nov.14 2020	& 0 &	remote	& 3.5151-4.6541\\ 
J072011-675631	&	4.70	& 	-/-/18.0	 	&	2.7	&	19/24	&	Jan.09 2022	& 55 &	remote	& 3.4110-4.5238\\ 
J120523-074232	&	4.69	& 	-/16.8/-	 	&	3	&	32/37	&	Apr.27-28 2019	& 29 &	on site	& 3.4680-4.5951\\ 
J221111-330245	&	4.64	& 	-/-/18.1	 	&	1.3	&	14/17	&	Sep.25 2021	& 76 &	remote	& 3.4220-4.5375\\ 
J143352+022713	&	4.62    & 	18.3/17.3/-	    &	3	&	25/29	&	Apr.28 2019	& 29 &	on site	& 3.4969-4.6314\\ 
J001225-484829	&	4.59	& 	-/-/17.5	 	&	3.0	&	33/38	&	Sep.25 2021	& 76 &	remote	& 3.4103-4.5230\\ 
J222152-182602	&	4.52	& 	17.8/-/-	 	&	2.5	&	28/33	&	Aug.16-17 2021 & 68 &	remote	& 3.3315-4.4244\\ 
J211920-772252	&	4.52	& 	-/-/17.4	 	&	2.3	&	20/24	&	Aug.16-17 2021 & 68 &	remote	& 3.3645-4.4656\\  
J233505-590103	&	4.50	& 	-/-/17.3	 	&	3.0	&	33/38	&	Aug.17 2021	& 78 &	remote & 3.3341-4.4275\\ 
J090634+023433	&	4.50	& 	18.5/17.2/-	    &	3.5	&	22/26	&	Dec.27 2019	& 4 &	on site	& 3.2971-4.3812\\ 
J040914-275632	&	4.45	& 	-/-/17.6	 	&	1.3	&	20/23	&	Sep.25 2021	& 76 &	remote	& 3.2902-4.3725\\ 	
J140801-275820	&	4.44	& 	17.8/-/-	 	&	1.4	&	15/18	&	Aug.16-17 2021 & 68 &	remote	& 3.2717-4.3493\\ 
J111700-111930	&	4.40	& 	18.8/17.0/-	    &	2.3	&	16/21	&	Dec.28 2019	& 9 &	on site	& 3.1884-4.2450\\ 
J000736-570151	&	4.25	& 	-/-/17.0	 	&	2.4	&	34/40	&	Aug.16-17 2021 & 68 &	remote	& 3.1225-4.1625\\
J101529-121314	&	4.19	& 	17.2/-/-	 	&	2.5	&	38/43	&	Mar.01 2022	& 0 &	remote	& 3.0615-4.0861\\ 	
J100114+211514	&	3.96	& 	18.4/16.5/-	    &	2.4	&	9/12	&	Jan.09 2022	& 55 &	remote	& 2.8947-3.8773\\  
\hline\hline
\end{tabular}\label{table:MIKEsample}
\end{center}
$\rm R_{p}$ is the Gaia red magnitude measured in $6400-10500\rm~\text{\AA}$ \citep{Wolf20}. $\rm S/N_{25\%}$ and $\rm S/N_{75\%}$ are 25th and 75th percentile $\rm S/N$ of pixels within the \ion{C}{4} absorber searching band (see an example in Fig.~\ref{fig:ExampleMIKESpec}). The last column $\Delta z_{\rm C{\scriptscriptstyle~IV}}$ is the corresponding redshift range of these searching bands. $P_{\rm moon}$ is the approximate moon phase (moon illumination percent) on the observation date(s).
\end{table*}

\subsection{Magellan/MagE Medium-Resolution Spectroscopy Observations} \label{subsec:MagellanMagE}

Most of the quasars at $z\gtrsim4$ are not bright enough (e.g., $\rm mag_i\lesssim19.0$) for the high resolution spectroscopy observation with MIKE (\S\ref{subsec:MagellanMIKE}). As a result, the numbers of quasar sightlines and high column density \ion{C}{4} absorbers in our MIKE sample are relatively small. Furthermore, our MIKE sample only covers the quasar redshift range of $z\lesssim5.2$ (Table~\ref{table:MIKEsample}; Fig.~\ref{fig:sample}), which limits the exploration of the cosmic reionization history at earlier stages. We therefore need a larger sample of slightly fainter quasars extending to a broader redshift range. Starting in the 2022B semester, we changed our focus to the medium resolution spectroscopy with the MagE spectrograph. 

The MagE (Magellan Echellette) Spectrograph is a moderate-resolution optical echellette mounted on the Magellan~I Baade telescope \citep{Marshall08}. Compared to MIKE, MagE has a slightly broader wavelength range of $\approx3,100-10,000\rm~\text{\AA}$ covered on a single detector, as well as a higher overall throughput at $\lesssim7,500\rm~\text{\AA}$ (peak throughput of $\sim0.2$ at $\sim5,200\rm~\text{\AA}$). MagE has eight $10^{\prime\prime}$-long slits, with widths of 0.5, 0.7, 0.85, 1.0, 1.2, 1.5, 2.0, and 5.0~arcsec. MagE has only one 175~lines/mm grating. When the seeing is good enough, we often choose the $0.5^{\prime\prime}$ slit, which combined with the grating gives a spectral resolution of $R\sim7,000$. This spectral resolution is high enough to detect and resolve the strongest and most complicated absorbers.

Until October 2023, we have already spent six nights with Baade/MagE in three semesters (2022B, 2023A, 2023B; another five nights allocated in 2024A and 2024B) to observe 29 quasars in a redshift range of $z\sim4.6-5.1$ (not yet observed any higher redshift ones). Among these quasar spectra, 12 reach a $\rm S/N_{25\%} \gtrsim 15~pixel^{-1}$. The brief MagE observation log is presented in Table~\ref{table:MagEsample} and an example MagE spectrum is presented in Fig.~\ref{fig:ExampleMagESpec}. Our final plan is to observe $\sim10^2$ relatively faint quasars with MagE in the next $\sim2-3$ years. Further analysis of the MagE spectra will be published in follow-up papers.

\subsection{Comparison with other samples} \label{subsec:CompareOthers}

We herein provide a rough estimate of the \ion{C}{4} detection limit for our MIKE and MagE samples for comparison with other samples. We caution that obtaining an accurate detection limit requires careful data reduction and numerical simulations for incompleteness correction. These detailed analyses will be presented in \S\ref{sec:MIKEDateReduction}, HIERACHY~III, and other follow-up papers.

Throughout the HIERACHY program, we use the weaker $\lambda1550.8\rm~\text{\AA}$ line to define the detection limit of the \ion{C}{4} absorbers at redshift $z_{\rm abs}$ and wavelength $\lambda = 1550.8 \times (1+z_{\rm abs})$ at 3~$\sigma$ confidence level. We follow the method in \citet{Burchett15} to calculate the detection limit in the rest-frame equivalent width $REW_{1550.8,\rm lim}$:
\begin{equation}
REW_{1550.8,\rm lim}(\lambda) = \frac{3 \sigma_{REW(\lambda)}}{1+z_{\rm abs}},\\
\end{equation}
where $\sigma_{REW(\lambda)}$ is the uncertainty of the observed equivalent width summed over a number of pixels, which is defined as:
\begin{equation}
\sigma_{REW(\lambda)}^2 = \sum_i \left( \Delta \lambda(i) \left[\frac{\sigma_{I(\lambda_i)}}{I(\lambda_i)} \right] \right)^2
\end{equation}
where $\Delta \lambda(i)$, $I(\lambda_i)$ and $\sigma_{I(\lambda_i)}$ are the wavelength range covered by a single pixel in $\text{\AA}$, continuum flux and flux uncertainty at pixel $i$, respectively. We use the number of pixels in the resolution element to determine the number of integrated pixels, i.e., $n_{\rm pix} = 2*\lambda_{\rm FWHM}/\Delta \lambda$ and typically range from four to eight. Here the factor of 2 is to ensure the integration region can cover most of the absorption profiles. We then convert $REW_{1550.8,\rm lim}$ to the corresponding column density detection limit with the linear part of the curve-of-growth: \begin{equation}
\frac{N_{\rm CIV,lim}}{10^{13} \rm~cm^{-2}} = \frac{REW_{1550.8,\rm lim}}{0.02 \rm~\text{\AA}}.
\end{equation}

\begin{figure}[!th]
\begin{center}
\includegraphics[width=0.55\textwidth]{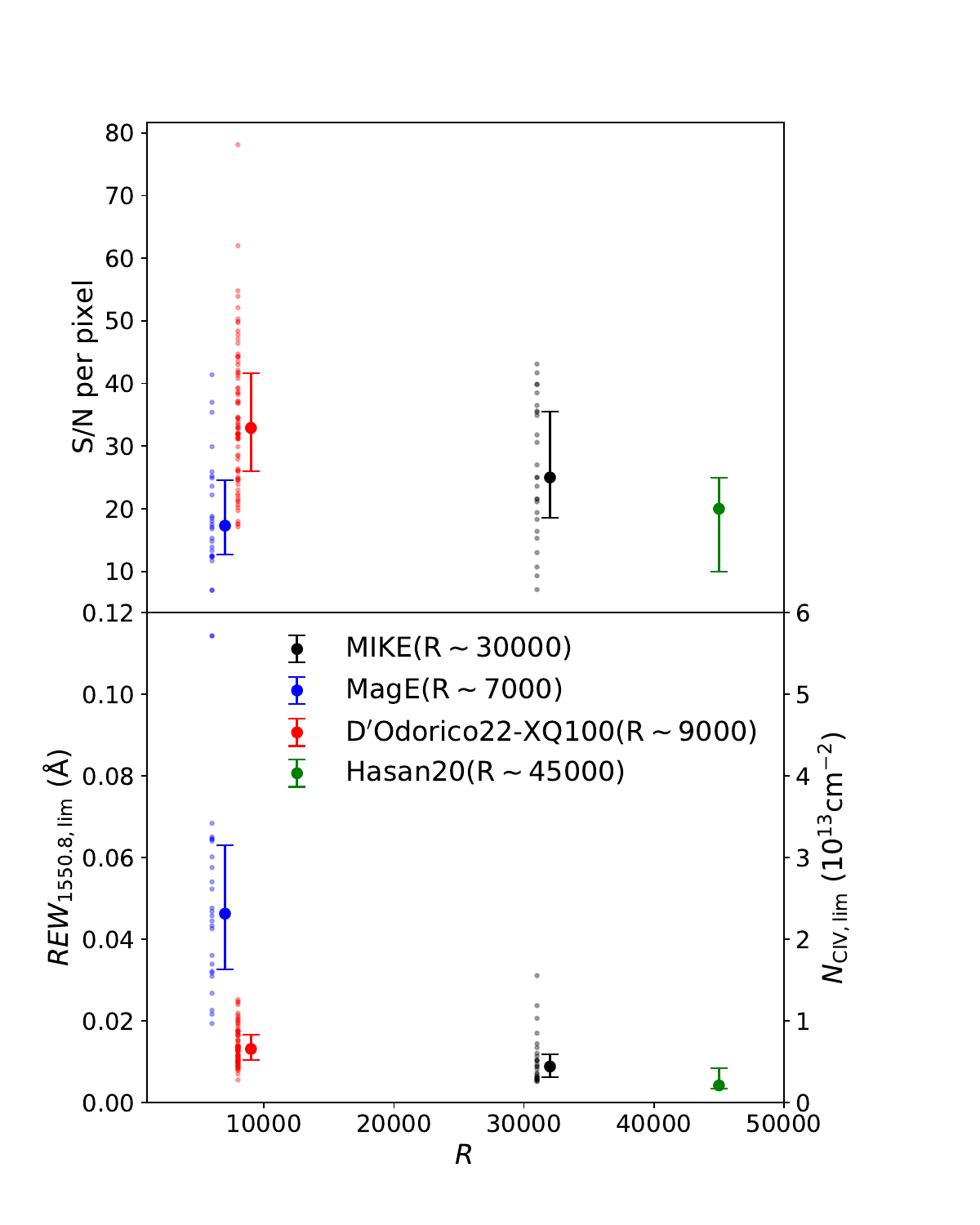}
\caption{Comparing the HIERACHY sample to the other similar samples on the data quality. The upper panel shows the spectral resolution $R$ and $S/N$ and bottom panel shows the estimated detection limits in the rest-frame equivalent width $REW_{1550.8,\rm lim}$ and column density $N_{\rm CIV,lim}$. Tiny dots are the median values of the $S/N$, $REW_{1550.8,\rm lim}$, or $N_{\rm CIV,lim}$ for all the pixels or calculation window in the \ion{C}{4} search region for individual quasar spectrum. Filled circles are the median value of the corresponding parameters of each sample, while the error bars are the 25th and 75th percentile value of the parameter's PDF. Note that we calculated the parameters of the XQ-100 sample \citep{DOdorico22} in the same way as our own MIKE and MagE data, but for \citet{Hasan20}'s sample, we directly quoted the parameters in their paper, which has been defined in a slightly different way.}\label{fig:SNRR}
\end{center}
\end{figure}

We compare the resolution, S/N, and detection limit of the spectra from our MIKE and MagE samples to some similar medium to high resolution spectroscopy observations of high-$z$ quasars \citep{Hasan20,DOdorico22} in Fig.~\ref{fig:SNRR}. Our MIKE sample has comparable detection limits to the best archival samples covering higher or lower redshift ranges. The MagE sample will be mostly useful to enlarge the high column density absorber sample with $N_{\rm CIV}>10^{13} \rm~cm^{-2}$, which is still useful for IGM study. Combining the two samples, the HIERACHY program provides the largest \ion{C}{4} searching range in the \ion{He}{2} reionization epoch (Fig.~\ref{fig:sample}), so could potentially provide the largest \ion{C}{4} absorber catalogue (HIERACHY~III).

\begin{table*}
\begin{center}
\caption{Bright quasars at $z\gtrsim4.5$ already observed with MagE until 2023B} 
\begin{tabular}{lcccccccc}
\hline\hline
 Quasar & $z$ & $\rm mag_{i}$ & $t_{\rm exp}$ & $\rm S/N~pixel^{-1}$ & Date & $P_{\rm moon}$ & Mode  & $\Delta z_{\rm C{\scriptscriptstyle~IV}}$  \\
 & & & (hr) & $25\%/75\%$ & & (\%) & &\\ 
\hline 
J065330+152604	&      4.90	& 	19.68	&	1.25	&	6/11	&	Dec.3 2022  &	85      &   remote	 &   3.6328-4.8017\\ 
\hline 
J075332+101511	&      4.89	& 	19.44	&	1.50	&	10/14	&	Dec.3 2022	&	85      &   remote	 &   3.6250-4.7918\\ 
\hline 
J024643+061045	&      4.57	& 	19.24 	&	1.50	&	11/13	&	Dec.3 2022	&	85      &   remote	 &   3.3737-4.4772\\ 
\hline 
J045427-050049	&      4.93	& 	18.80 	&	1.64	&	21/31	&	Feb.14 2023	&  	33      &   remote   &   3.5403-4.6854\\ 
\hline 
J044432-292419	&      4.80	& 	18.40	&	1.42	&	8/16	&	Feb.14 2023	&	33      &   remote   &   3.5543-4.7030\\ 
\hline 
J095139-274212	&      4.80	& 	18.53	&	1.50	&	25/36	&	Feb.15 2023	&	23      &   remote    &  3.5543-4.7030\\ 
\hline 
J110837-185408	&      4.79	& 	18.96	&	1.49	&	14/22	&	Feb.14 2023	&	33      &   remote	 &   3.5335-4.6769\\ 
\hline 
J123141-184149	&      4.79	& 	18.98	&	1.33	&	13/21	&	Feb.14 2023	&	33     &   remote	 &   3.5467-4.6935\\  
\hline 
J094409+100656  &      4.77	& 	19.21	&	1.36	&	10/18	&	Feb.15 2023	&	23     &   remote	 &   3.5240-4.6651\\  
\hline 
J045057-265541	&      4.76	& 	18.74	&	1.50	&	18/27	&	Feb.15 2023	&	23      &   remote	 &   3.5229-4.6636\\ 
\hline 
J121402-123548	&      4.74	& 	18.36	&	1.50	&	18/29	&	Feb.15 2023	&	23      &   remote	 &   3.5075-4.6443\\ 
\hline 
J125049-065758	&      4.73	& 	18.70	&	1.76	&	17/31	&	Feb.15 2023	&	23      &   remote	 &   3.4996-4.6345\\ 
\hline 
J130031-282931	&      4.69	& 	18.00	&	1.50	&	29/41	&	Feb.14 2023	&	33      &   remote	 &   3.4682-4.5952\\ 
\hline 
J203310+121851	&      5.11	& 	19.11	&	2.74	&	11/15	&	Aug.03 2023	&	91      &   remote	 &   3.7978-5.0079\\ 
\hline 
J142721-050353	&      5.08	& 	19.41	&	1.50	&	16/21	&	Aug.03 2023	&	91      &   remote	 &   3.7742-4.9784\\ 
\hline 
J153359-181027	&      5.06	& 	19.38	&	1.82	&	12/18	&	Aug.03 2023	&	91      &   remote	 &   3.7271-4.9189\\ 
\hline 
J151911-065042	&      5.03	& 	19.51	&	1.83	&	10/17	&	Aug.04 2023	&	83      &   remote	 &   3.7059-4.8928\\
\hline 
J013539-212628	&      4.94 & 	18.13	&	1.50	&	33/42	&	Aug.03 2023	&	91      &   remote	 &   3.6645-4.8410\\
\hline 
J222357-252634	&      4.80 & 	18.68	&	1.50	&	15/20	&	Aug.03 2023	&	 91     &   remote	 &   3.5524-4.7006\\
\hline 
J215216+104052	&      4.79 & 	18.39	&	1.50	&	12/18	&	Aug.04 2023	&	 83     &   remote	 &   3.5543-4.7030\\
\hline 
J005527+122840	&      4.70 & 	18.87	&	1.50	&	11/16	&	Aug.04 2023	&	 83     &   remote	 &   3.4775-4.6068\\
\hline 
J220158-202627	&      4.67 &   18.27	&	1.50	&	22/28	&	Aug.03 2023	&	 91     &   remote	 &   3.5075-4.6443\\
\hline 
J201939-194717	&      4.61 &   18.61	&	1.50	&	15/23	&	Aug.04 2023	&	 83     &   remote	 &   3.3954-4.5041\\
\hline 
J215728-360218	&      4.67 &   17.37	&	1.50	&	36/50	&	Aug.04 2023	&	 83     &   remote	 &   3.5263-4.6674\\
\hline 
J024601+035054	&      4.96 & 	19.48	&	1.50	&	5/9     &	Oct.24 2023	&	 83     &   remote	 &   3.6803-4.8602\\
\hline 
J015618-044139	&      4.94 & 	19.21	&	1.50	&	10/14     &	Oct.24 2023	&	 83     &   remote	 &   3.6646-4.8406\\
\hline 
J225944+093624	&      4.86 & 	18.93	&	1.70	&	10/15     &	Oct.24 2023	&	 83     &   remote	 &   3.6014-4.7620\\
\hline 
J035954+054420	&      4.82 & 	18.53	&	1.50	&	15/21     &	Oct.24 2023	&	 83     &   remote	 &   3.5702-4.7229\\
\hline 
J205724-003018	&      4.68 & 	18.77	&	1.31	&	5/10     &	Oct.24 2023	&	 83     &   remote	 &   3.4601-4.5850\\
\hline\hline
\end{tabular}\label{table:MagEsample}
\end{center}
\end{table*}

\begin{figure*}[!th]
\begin{center}
\includegraphics[width=1.0\textwidth]{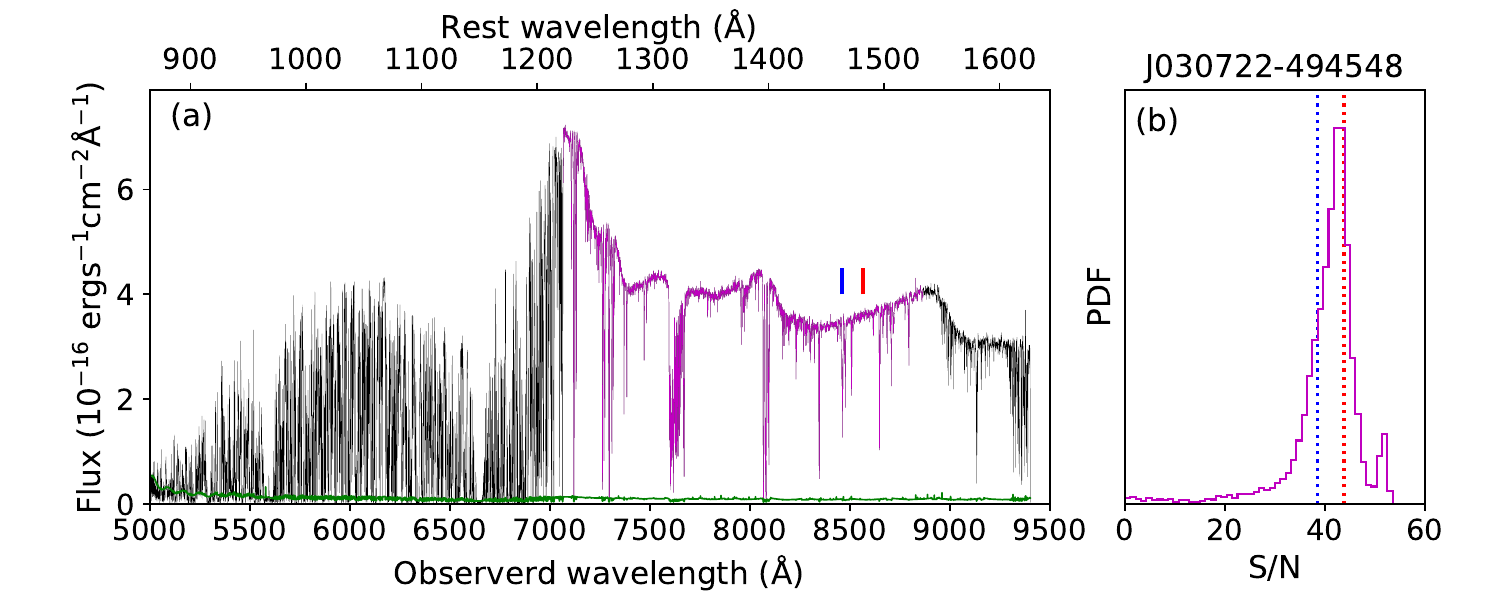}
\includegraphics[width=0.47\textwidth]{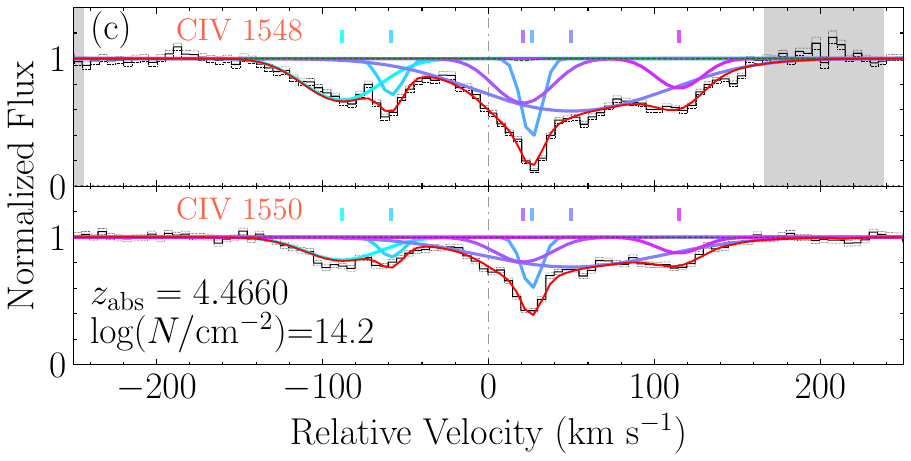}
\includegraphics[width=0.49\textwidth]{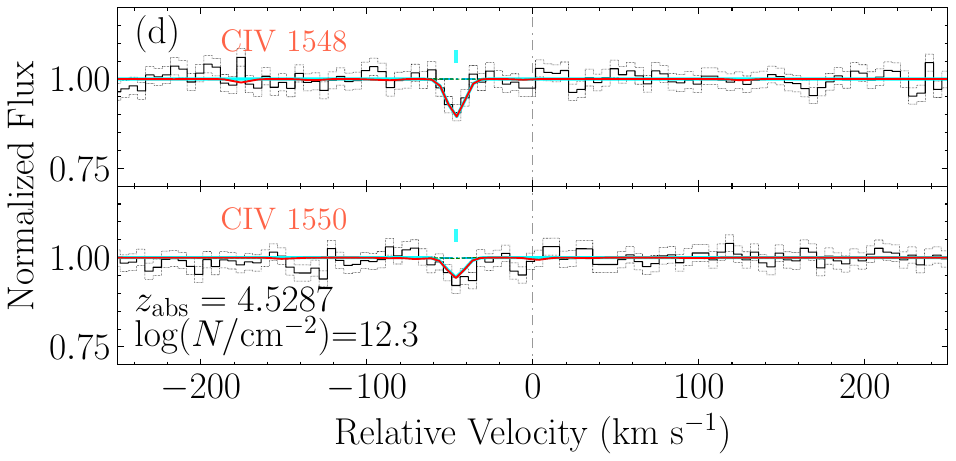}
\caption{An example Magellan/MIKE red channel spectrum of the $z=4.78$ quasar J030722-494548. (a) is the smoothed global spectrum in the full band of the red channel, with the \ion{C}{4} searching region highlighted in magenta. The green curve at the bottom shows the flux error. (b) shows the probability distribution function (PDF) of the $S/N$ of each pixel within the \ion{C}{4} searching region. The two vertical dotted lines mark the 25th (blue) and 75th (red) percentile $\rm S/N$ ($\rm S/N_{25\%}$, $\rm S/N_{75\%}$; see definition in the text). The short blue and red bars in panel (a) mark the location of two representative (strong and weak) identified \ion{C}{4} absorption systems, the $1548\rm~\text{\AA}$ and $1550\rm~\text{\AA}$ doublets of which are further zoomed-in in (c) and (d), respectively. The colored bars in (c) and (d) marks the centroid velocity of each identified \ion{C}{4} absorption components, while the grey shaded area is the band highly contaminated by either the sky emission lines or absorption lines from other ions.}\label{fig:ExampleMIKESpec}
\end{center}
\end{figure*}

\begin{figure*}[!th]
\begin{center}
\includegraphics[width=1.0\textwidth]{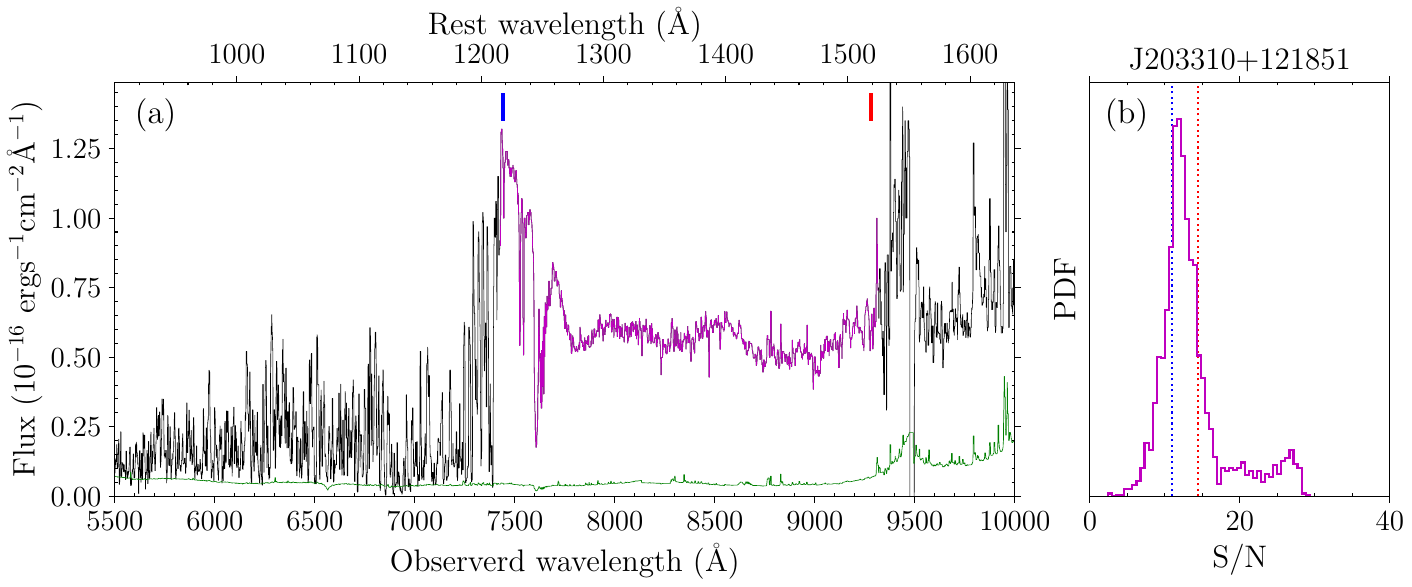}
\includegraphics[width=0.49\textwidth]{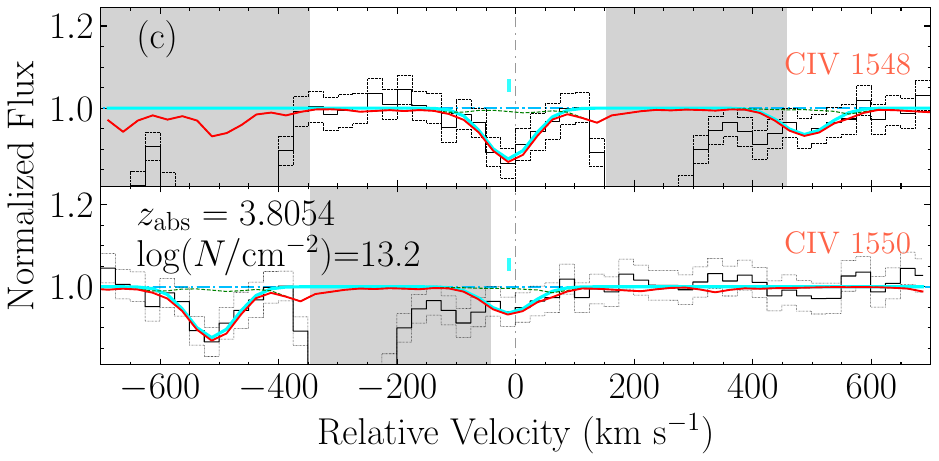}
\includegraphics[width=0.49\textwidth]{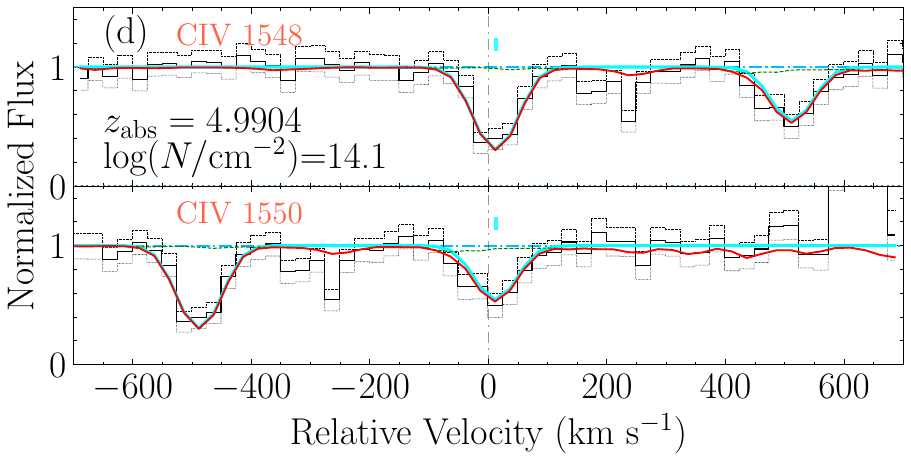}
\caption{Similar as Fig.~\ref{fig:ExampleMIKESpec}, but for an example of the Magellan/MagE spectrum of J203310+121851 at $z=5.11$. The flux gap at $\lambda\sim9500\rm~\text{\AA}$ is from the spectral order edges. The $z\approx4.99$ absorber is one of the highest redshift \ion{C}{4} absorbers identified in the HIERACHY program.}\label{fig:ExampleMagESpec}
\end{center}
\end{figure*}

\section{Magellan/MIKE data reduction procedure} \label{sec:MIKEDateReduction}

In this section, we introduce our data reduction procedures for the high-resolution Magellan/MIKE spectra. Further analysis of the data, including the estimation of the detection limit and \ion{C}{4} absorber sample completeness, the definition of different types of absorbers, and the construction of the \ion{C}{4} absorber catalogue, will be presented in HIERACHY~III. The analysis procedure for the Magellan/MagE data is quite similar; however, since the MagE observations are still ongoing, we will present detailed reduction and analysis procedures for the MagE data in follow-up papers. In this paper, we only present an example of the reduced MagE spectra in Fig.~\ref{fig:ExampleMagESpec}.

The MIKE raw data is reduced using the MIKE-dedicated pipeline, CarPy \citep{Kelson00,Kelson03}\footnote{ https://code.obs.carnegiescience.edu/mike}, 
which includes overscan subtraction, pixel-to-pixel flat field correction, image coaddition, cosmic ray removal, sky and scattered-light subtraction, rectification of the tilted slit profiles along the orders, spectrum extraction, and wavelength calibration. We use the default CarPy settings to extract the spectra, which are sufficient for standard observations of bright point-like sources.

Further calibrations of the extracted quasar spectra follow the procedures outlined in HIERACHY~I and the associated erratum \citep{Yu23}, including flux calibration, heliocentric velocity correction, and air-to-vacuum wavelength correction. For flux calibration, we use the ESO X-shooter standard star reference spectra\footnote{https://www.eso.org/sci/observing/tools/standards/spectra/Xshooterspec.html}. The heliocentric velocity $V_{\rm h}$ is estimated using the Python package \texttt{helcorr} and then applied to the original spectra. Additionally, the air-to-vacuum wavelength correction is performed using the Python package \texttt{airtovac2}.

\begin{figure*}
\begin{center}
\includegraphics[width=1\textwidth]{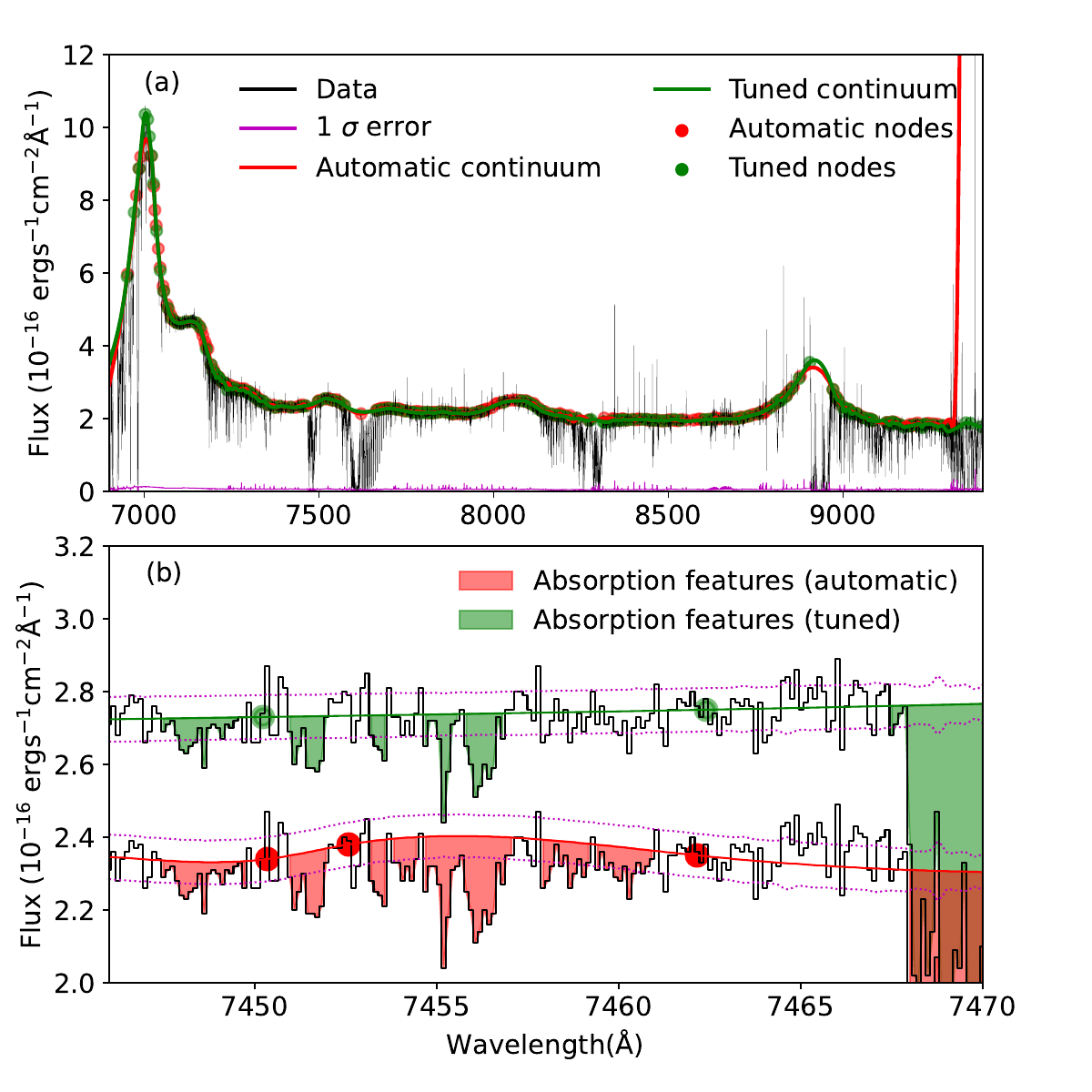}
\caption{An example spectrum showing the method determining the global continuum. The black and magenta curves show the original data and its 1~$\sigma$ error. The red curve and filled circles are the automatically determined global continuum and the continuum nodes, using the method detailed in \S\ref{sec:MIKEDateReduction}. The green filled circles are the manually adjusted continuum nodes, while the green curve is the final global continuum determined from a spline fit to these nodes. Panel (b) is a zoom-in of panel (a), and shows the effect of adjusting the continuum nodes to obtain a more reasonable global continuum. The corresponding absorption features are shown in red and green shadows.} 
\label{fig:ctngb}
\end{center}
\end{figure*}

\begin{figure*}
\begin{center}
\includegraphics[width=1\textwidth]{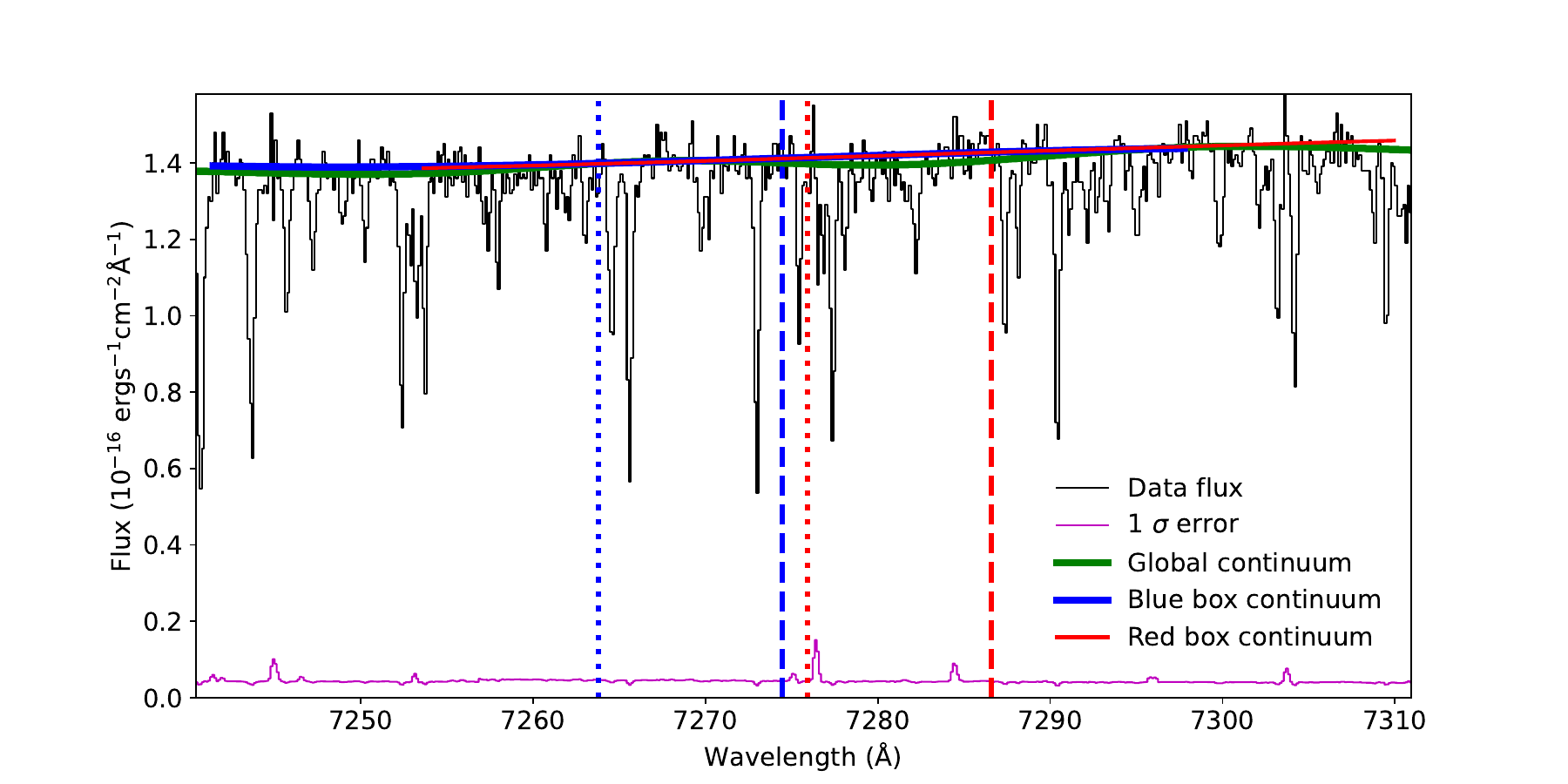}
\caption{An example showing the determination of the search window and the local continuum, as detailed in \S\ref{sec:MIKEDateReduction}. The blue and red vertical lines shows the search windows of \ion{C}{4} $\lambda1548$\text{\AA} and $\lambda1550$\text{\AA}, respectively. The green solid curve shows the initial global continuum, while the blue and red curves show the redefined local continuum for the ion{C}{4} $\lambda1548$\text{\AA} and $\lambda1550$\text{\AA} search windows, respectively.} 
\label{fig:ctnloc}
\end{center}
\end{figure*}

Our primary goal is to identify and measure various absorption lines in the quasar spectra. For these purposes, an accuracy estimate of the global continuum is typically not critical, especially for absorption lines located in the wavelength range longer than the Ly$\alpha$ emission line of the quasar in the observed frame. To facilitate absorption line identification, we first classify the pixels in the quasar spectra into three categories: absorption features, emission features, and continuum. This classification helps in defining the local continuum necessary for accurate absorption line identification. 

We begin by making an initial guess of the continuum by calculating the median flux value within a 401-pixel ($\Delta v=2005\rm~km~s^{-1}$) window centered at each spectral pixel (using the reflection of neighboring pixels for those at the edges of the spectra). Since the continuum can be significantly affected by strong and broad absorption features produced either by intervening gas or the Earth's atmosphere, it is necessary to mask the pixels heavily influenced by these absorption features. These features are identified using the following criterion: $(D_i - C_i)/E_i > -1.5$, where $D$, $E$, $C$ represent the flux, flux error, and estimated global continuum calculated with the window above for pixel $i$, respectively. This process is repeated twice to achieve a smoother global continuum. We then perform a third smoothing of the spectra using a smaller 201-pixel window. Absorption and emission features are defined as those having fluxes smaller (for absorption) or larger (for emission) than the global continuum with a significance greater than 3~$\sigma$ and 5~$\sigma$, respectively. Here, the significance of the feature is calculated over at least three pixels and is defined as: $\sigma=\sqrt{\sum_{i} (\frac{D_i-C_i}{E_i})^{2}}$. A stricter criterion is applied to emission features because some broad emission lines in the quasar spectra can be used as the continuum when searching for absorption lines.

After these three iterations, we typically achieve an adequate global continuum, shown as the red curve in Fig.~\ref{fig:ctngb}. We then define nodes for each continuum feature between neighboring emission or absorption features by calculating the median flux covered by a continuous continuum feature. These nodes are represented as red circles in Fig.~\ref{fig:ctngb}. This initial continuum guess needs further adjustment to mitigate the impact of strong fluctuations, such as those caused by residual weak emission or absorption features. This adjustment is conducted through a visual check, where we remove nodes clearly affected by non-continuum features and manually add nodes at ``clean'' pixels (green circles in Fig.~\ref{fig:ctngb}). Finally, we perform a spline fit to these manually adjusted nodes to determine the global continuum and then use the same criteria as above to define the emission, absorption, or continuum features.

We then perform an automated search for various absorption lines using a moving search window with Voigt profile fittings. Before initiating the search, we exclude pixels that are heavily contaminated by known telluric lines or strong absorption features from other ions, as identified in earlier iterative analysis steps. This exclusion process is conducted separately for each ion. The specific criteria for identifying these contaminated pixels, along with additional inspection procedures, will be detailed in HIERACHY~III. As an example, we focus here on the \ion{C}{4} doublet. In our scientific analysis of intervening \ion{C}{4} absorbers, we will only include \ion{C}{4} absorbers located at velocities of $\leq-5000\rm~km~s^{-1}$ relative to the \ion{C}{4} emission line of the quasar. However, in this step, we do not impose an upper wavelength limit on the search window, allowing us to also capture \ion{C}{4} absorption lines potentially associated with quasar outflow or inflow phenomena.

We first need to define the search window for absorption lines within narrow wavelength ranges, aiming to more accurately determine the local continuum. The initial analysis above provides the wavelength range of different absorption, emission, and continuum features, as well as an initial global continuum. We set the blue end of a search window at the blue edge of an absorption feature. The width of the search window is initially set to be approximately $450\rm~km~s^{-1}$. If the red end of the search window reaches another absorption feature, either its own or a different one, we adjust the red end of the search window to coincide with the red edge of that absorption feature. The search window is then moved across the full absorption line search range, which typically extends from the quasar's Ly$\alpha$ emission line to the red end of the MIKE spectra. An example of such search windows for the \ion{C}{4} $\lambda\lambda1548,1550\rm~\text{\AA}$ doublet is shown in Fig.~\ref{fig:ctnloc}.

In addition to the absorption line search windows, we also define windows for continuum features. These windows are positioned adjacent to the absorption line search window and extend at least $200\rm~km~s^{-1}$ from both ends. We then search for pixels identified as continuum features within these windows and use a spline fit to these pixels to determine the local continuum. This local continuum is often slightly different from the global continuum defined earlier and is more accurate for a narrow wavelength range (Fig. \ref{fig:ctnloc}). If no continuum feature is present in the vicinity of the absorption line search window --- such as when the search window falls within broad absorption lines (BALs) of quasar outflows or broad telluric absorption features --- we use the global continuum determined in the previous step instead. Finally, we renormalize the local continuum to unity to calculate the equivalent width of each identified absorption line.

We adopt the TAPAS telluric template to model the telluric absorption features \citep{Bertaux14}. This template includes absorption lines from various molecules, such as $\rm O_2$, $\rm H_2O$, $\rm O_3$, $\rm CO_2$, $\rm CH_4$, and $\rm N_2O$. However, only $\rm O_2$ and $\rm H_2O$ are included in our analysis because the other lines are not resolvable within our wavelength range of interest. We estimate the telluric absorption strength using the transmission template $T(v)$, adjusted with an index $a_{\rm T}$, which is a free parameter determined during the spectrum fitting.

We use the python package lmfit to fit the \ion{C}{4} doublet with Voigt profiles. Each of the Voigt component includes three free parameters: the velocity of the line center $v_{\rm c}$, the Doppler parameter $b$ describing the broadening of the line, and the column density $\log N$. The initial value of $v_{\rm c}$ is set at the pixel with the deepest absorption. If the initial fit is not satisfactory, we add another Voigt profile until the significance of this additional component is smaller than 3~$\sigma$ --- that is, when $\Delta\chi_{\rm before}$ - $\Delta\chi_{\rm after} < 15.63$.

Some examples of the identified \ion{C}{4} and other absorption lines from the MIKE spectra are presented in Fig.~\ref{fig:ExampleMIKESpec} and other figures in subsequent sections. Further discussions on contamination, detection limits, and additional examples of detected \ion{C}{4} lines at different confidence levels will be presented in HIERACHY~III.

\section{Major Scientific Goals and Examples of Initial Data Products} \label{sec:SciGoals}

The major scientific goals of the HIERACHY program are to study the properties of the IGM and the reionization processes during the \ion{He}{2} EoR. These studies are primarily based on quasar absorption line analyses (\S\ref{subsec:SciGoalzEvolvNCIVFunction}, \ref{subsec:SciGoalMultiphaseIGM}, \ref{subsec:SciGoalLyaForestIGMEoS}). Additionally, we aim to investigate related processes occurring during a similar stage of cosmic evolution, such as galaxies and extended nebulae (\S\ref{subsubsec:SciGoalCIVHostGalaxy}), large-scale structures (\S\ref{subsubsec:SciGoalProtoCluster}), and AGN outflows (\S\ref{subsubsec:SciGoalOutflowInflow}). These additional scientific goals are also closely tied to the \ion{He}{2} reionization history and are explored using both our quasar absorption line data and follow-up multi-wavelength observations. In the following sub-sections, we will briefly introduce these scientific goals, along with some initial examples of results from the direct data products.


\subsection{Statistical properties and redshift evolution of \ion{C}{4} absorbers}\label{subsec:SciGoalzEvolvNCIVFunction}

The \ion{C}{4} doublet is the easiest to detect in optical spectra at the redshift of interest. Moreover, due to the comparable ionization potentials of the C$^{3+}$ and He$^+$ ions (64.5~eV vs 54.4~eV), the \ion{C}{4} doublet serves as an effective tracer of the \ion{He}{2} reionization. Our first major scientific goal is therefore focused on the statistical properties and redshift evolution of \ion{C}{4} absorbers, using them to track the \ion{He}{2} reionization history. 

The comoving mass density of C$^{3+}$ ions ($\Omega_{\rm C{\scriptscriptstyle~IV}}$) derived from high column density ($\log N_{\rm C{\scriptscriptstyle~IV}}/\rm cm^{-2}\gtrsim13.0$) \ion{C}{4} absorbers does not show significant redshift evolution in a broad redshift range at $z\lesssim5$ (or slight increase toward low-$z$; e.g., \citealt{Songaila01,Songaila05,Pettini03,Cooksey10,Cooksey13,DOdorico10,Shull14}), but could change significantly at higher redshifts (e.g., \citealt{Songaila05,Becker09,RyanWeber09,DOdorico13,Codoreanu18,Cooper19,Davies23b}). This suggests that \ion{He}{2} reionization was completed at $z\gtrsim5$, which conflicts with the known cosmic reionization history (e.g., \citealt{Davidsen96,McQuinn09,FaucherGiguere09,McQuinn16}), unless significant carbon enrichment occurred early at $z\gtrsim5$. However, many existing measurements of the C$^{3+}$ comoving mass density are primarily based on the high column density absorbers with $\log N_{\rm C{\scriptscriptstyle~IV}}/\rm cm^{-2}\gtrsim13.0$. These absorbers may be highly biased by outflows from star formation or AGN-driven winds, or by the CGM in the immediate vicinity of galaxies, and therefore may not accurately reflect the true IGM evolution (e.g., \citealt{Pettini03}). Probing the \ion{C}{4} absorbers down to a lower detection limit (e.g., to $\log N_{\rm C{\scriptscriptstyle~IV}}/\rm cm^{-2}\sim12.0$) is thus crucial for understanding the properties of C$^{3+}$ gas in the genuine IGM (e.g., \citealt{Ellison00,DOdorico10,DOdorico16,Kim13}). 

With our large sample of high resolution, high S/N spectra, we can detect \ion{C}{4} absorbers with column densities as low as $\log N_{\rm C{\scriptscriptstyle~IV}}/\rm cm^{-2}\sim12.0$ (e.g., Fig.~\ref{fig:ExampleMIKESpec}). This enables us to construct the \ion{C}{4} column density function $f(N_{\rm C{\scriptscriptstyle~IV}})$ at different redshift intervals down to a limit well below the threshold where contamination from the CGM is expected to be significant (e.g.,  \citealt{Ellison00,Songaila01}). This threshold is often around $\log N_{\rm C{\scriptscriptstyle~IV}}/\rm cm^{-2}\lesssim13-14$, below which the \ion{C}{4} column density profile around galaxies shows a clear steepening toward larger radii (e.g., \citealt{Steidel10}). We will further integrate $f(N_{\rm C{\scriptscriptstyle~IV}})$ to derive the C$^{3+}$ comoving mass density $\Omega_{\rm C{\scriptscriptstyle~IV}}$, separating the high column density CGM component from the low column density IGM component, and study its redshift evolution at $z\sim3-4.5$ (e.g., \citealt{Songaila01,Pettini03}).

In addition to the \ion{C}{4} column density function and the integrated \ion{C}{4} comoving mass density $\Omega_{\rm C{\scriptscriptstyle~IV}}$, there are other parameters based solely on \ion{C}{4} absorber measurements that can be used to probe the IGM during the \ion{He}{2} EoR. For example, the \ion{C}{4} absorber number density per comoving path-length $d\mathcal{N}_{\rm C{\scriptscriptstyle~IV}}/dX$ is often used to traced the redshift evolution of either the reionizaiton or carbon enrichment processes (e.g., \citealt{Cooksey10,Cooksey13,DOdorico10,DOdorico22,Davies23}). The two-point correlation function of the \ion{C}{4} forest can be employed to measure the metallicity of weak IGM absorbers (e.g., \citealt{Tie22}). The broadening of the weakest \ion{C}{4} absorbers could help constrain the IGM temperature (e.g., \citealt{Rauch97}). Moreover, with the assistance of Ly$\alpha$ forest analysis, stacking or optical depth analysis may also aid in measuring the IGM metallicity (\S\ref{subsec:SciGoalLyaForestIGMEoS}). Some of these analyses will be presented in HIERACHY~III, while others will be further explored in follow-up papers.


\subsection{The multi-phase CGM/IGM}\label{subsec:SciGoalMultiphaseIGM}

\begin{figure*}[htb]
\centering
\includegraphics[width=1.0\textwidth]{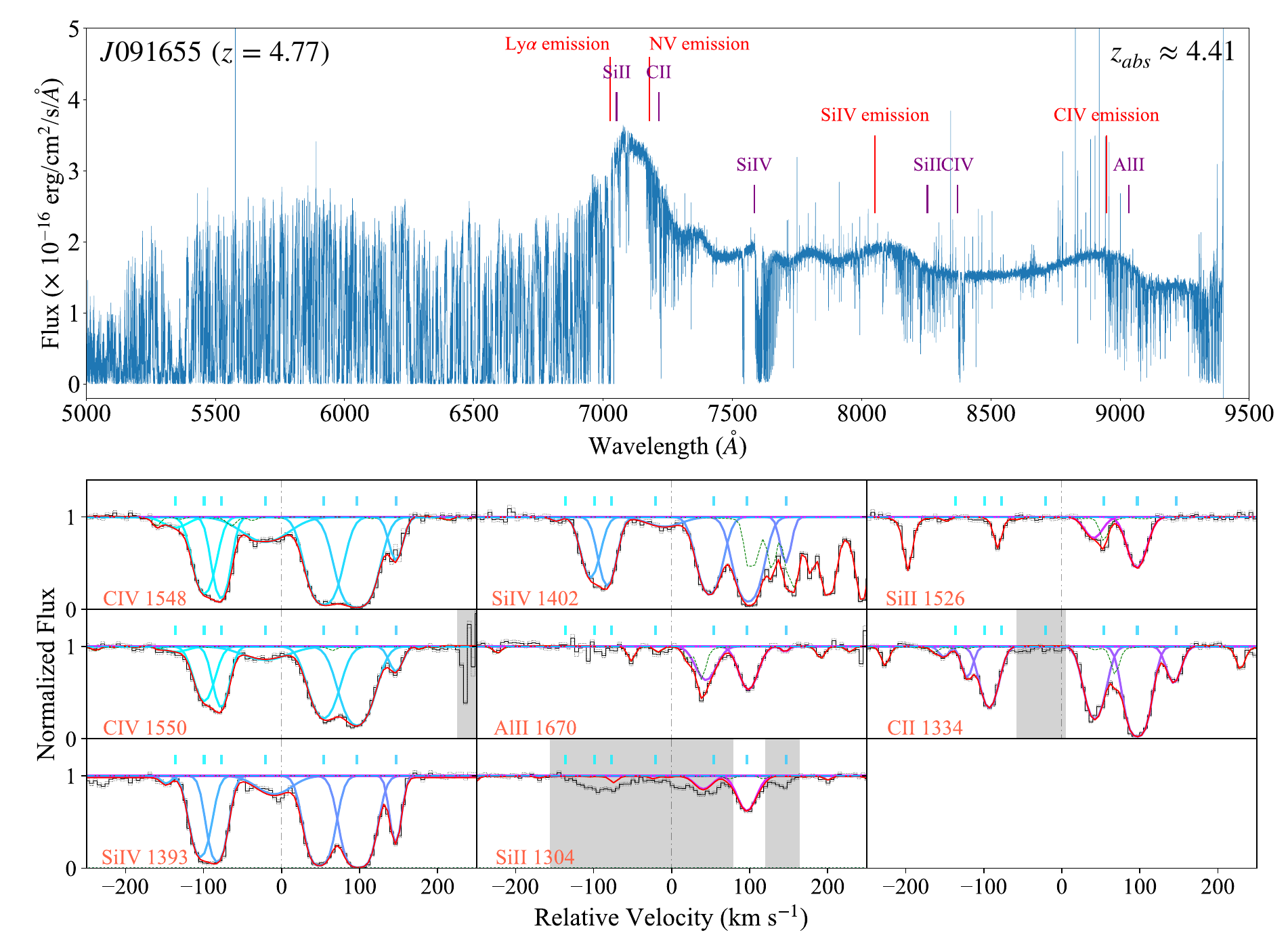}
\caption{An example of a $z\approx4.41$ multi-ion absorption system in our Magellan/MIKE spectrum of the background quasar J091655 at $z\approx4.77$. In addition to the strong \ion{C}{4} absorption doublets, we also identify coherent fine structures in the \ion{Si}{4}, \ion{Si}{2}, \ion{C}{2} and \ion{Al}{2} absorption components. The short colored bars mark the velocity centers of each \ion{C}{4} absorption components. The shaded areas mask bands clearly contaminated by either the telluric lines or absorption lines from other ions.}\label{fig:J091655}
\end{figure*}

\begin{figure*}[htb]
\centering
\includegraphics[width=1.0\textwidth]{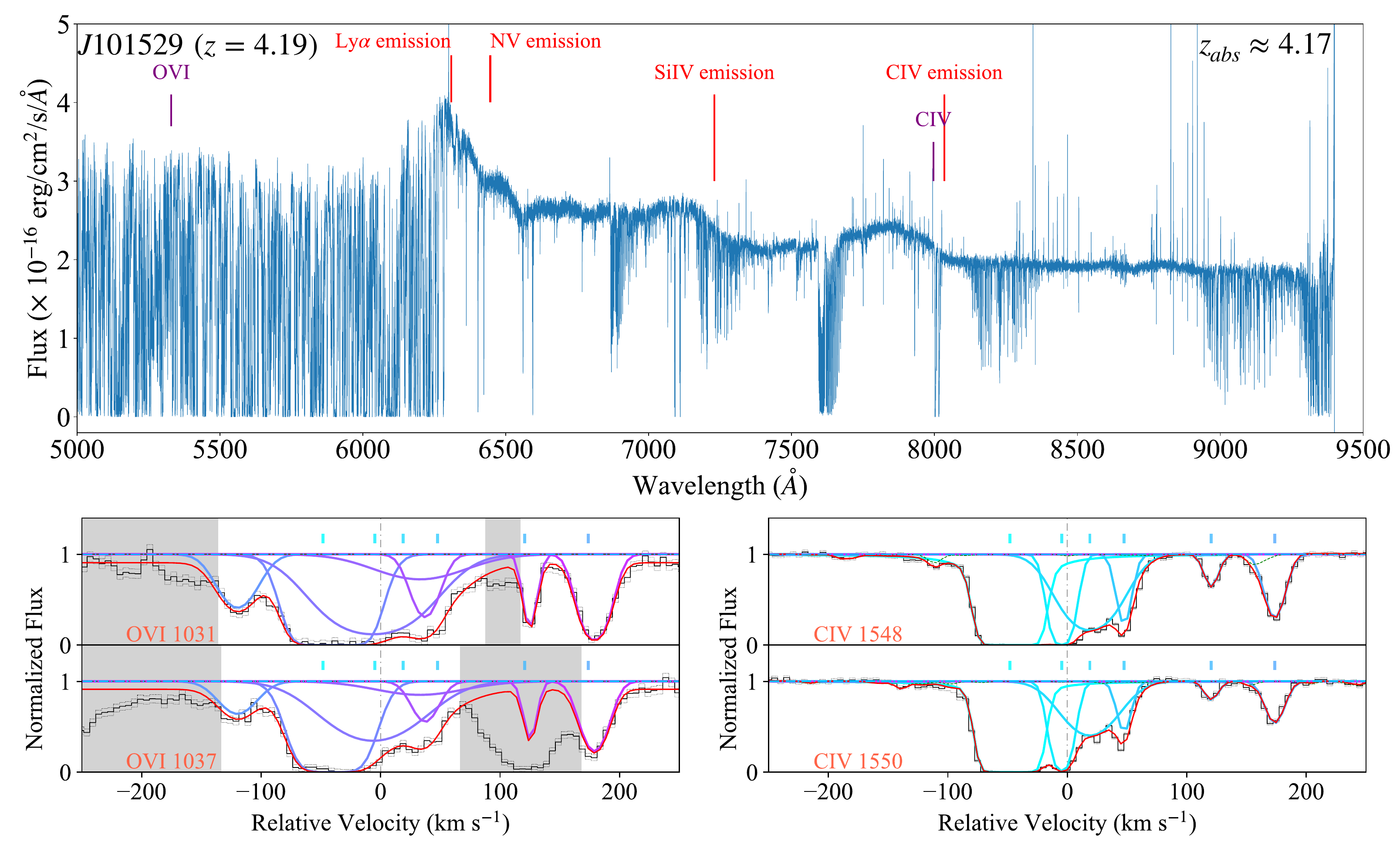}
\caption{An example of an \ion{O}{6} absorber detected likely in a strong quasar outflow as resolved in our Magellan/MIKE spectrum. The quasar J101529 is located at $z\approx4.19$, while the strong and broad \ion{C}{4} absorption system has a center redshift of $z\approx4.17$. The above panel shows the location of the \ion{C}{4} and \ion{O}{6} absorbers, while the lower zoom-in panels show individual components of them, which have coherent substructures. Symbols are the same as in Fig.~\ref{fig:J091655}.}\label{fig:OVI}
\end{figure*}

The multi-phase gases in the CGM/IGM at a temperature of $T\sim10^{4-6}\rm~K$ are characterized by various metal absorption lines in the rest frame UV band (e.g., \citealt{Rauch97,Werk13,Shull14,Turner16}). In our primary data analysis approach (partially presented in HIERACHY~I and to be detailed further in HIERACHY~III), we first identify the \ion{C}{4}~$\lambda\lambda1548,1550\rm~\text{\AA}\AA$ doublet, which are often relatively strong compared to other metal lines, and have well-defined wavelength and flux ratios (e.g., \citealt{Feibelman83,Petitjean04}). We then search for other metal lines from different ions at the same redshift as the identified \ion{C}{4} lines. This search is primarily conducted for lines at longer wavelengths than the Ly$\alpha$ emission line in the quasar spectrum (e.g., Fig.~\ref{fig:J091655}), though the strongest metal absorbers can still be identified even when embedded within the Ly$\alpha$ forest (e.g., Fig.~\ref{fig:OVI}). 

\begin{figure*}[!th]
\centering
\includegraphics[width=1.0\textwidth]{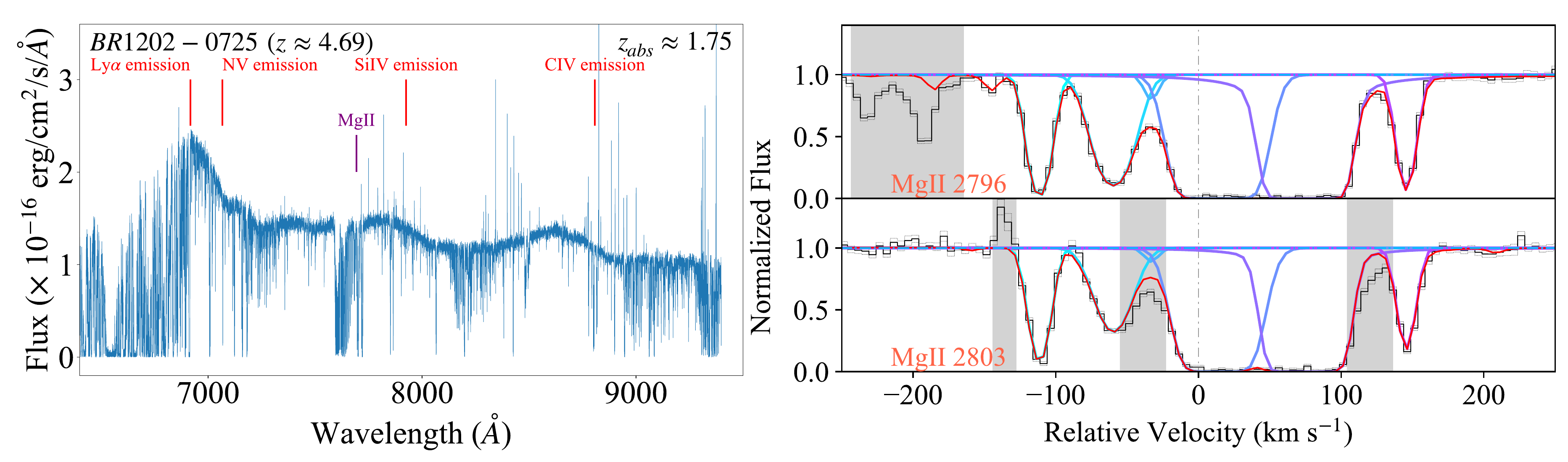}
\caption{An example of a strong \ion{Mg}{2} absorber located at $z\approx1.75$ in the spectrum of the background quasar BR~1202-0725 (J120523-074232 at $z\approx4.69$ in  Table~\ref{table:MIKEsample}; also see HIERACHY~I). Symbols are the same as in Figs.~\ref{fig:J091655}, \ref{fig:OVI}.}\label{fig:MgII}
\end{figure*}

Depending on the redshift of the absorber, there are several commonly detected metal lines redder than Ly$\alpha$, including \ion{C}{2}~$\lambda1334\rm~\text{\AA}$, \ion{N}{5}~$\lambda\lambda1239,1243\rm~\text{\AA}\AA$, \ion{Si}{4}~$\lambda\lambda1394,1403\rm~\text{\AA}\AA$, \ion{Fe}{2}~$\lambda1608\rm~\text{\AA}$, and, at significantly lower redshifts,
\ion{Mg}{2}~$\lambda\lambda2796,2803\rm~\text{\AA}\AA$ (Fig.~\ref{fig:MgII}). Among these, the \ion{Si}{4} and \ion{Mg}{2} (at lower redshifts) doublets are often the most common and strongest, sometimes even serving as indicators for identifying other metal absorbers. Whenever possible, we also search for important lines bluer than the Ly$\alpha$ emission line (e.g., \ion{O}{6}; \citealt{Simcoe04}). These lines are embedded within the Ly$\alpha$ forest, so identifying their weaker components requires detecting coherent absorption lines at longer wavelengths (e.g., \ion{C}{4}), carefully modeling the entire Ly$\alpha$ forest, and/or removing high-order Lyman series lines (Ly$\beta$ or higher).

The identified absorption lines from multiple ions with different ionization potentials are excellent tracers of the multi-phase IGM/CGM. Including both high and low ionization ions is crucial for distinguishing between different gas phases (e.g., \citealt{Rauch97}). We will examine various photo-ionization models using the Cloudy code \citep{Ferland13,Ferland17} to estimate numerous physical properties of the gas, such as metallicity and the ionization parameter (the $U$ factor; \citealt{Shull14}). In addition to modeling the multi-ion strong absorbers, we will also stack spectra around different absorption lines at the redshift of the identified weak \ion{C}{4} lines. This approach will help us study weaker multi-ion systems. 

We will then examine the redshift evolution of the physical parameters of these multi-ion systems to better understand the thermal and ionization history of the Universe during the \ion{He}{2} EoR. For instance, evidence suggests that changes in the \ion{Si}{4}/\ion{C}{4} line ratio at $z\sim3$ may indicate significant hardening of the UV background at the end of the \ion{He}{2} EoR (e.g., \citealt{Songaila98,Songaila05}), but such a change in the \ion{Si}{4}/\ion{C}{4} line ratio is not confirmed in other works (e.g., \citealt{Boksenberg15}). Moreover, by combining Ly$\alpha$ (\S\ref{subsec:SciGoalLyaForestIGMEoS}) and \ion{C}{4} (or other metal absorption lines) measurements for some absorbers, we will also assess the metallicity of different elements, construct their metallicity distribution functions, and derive the metal enriched IGM mass function (e.g., \citealt{Simcoe04,DOdorico22}). This will enable us to estimate the total metal content at specific redshift intervals. Furthermore, we will investigate the dependence of metallicity on the gas density or column density (e.g., \citealt{Rauch97}). This will also allow us to determine whether there is a universal metallicity floor at low densities, as predicted by some Population~III enrichment scenarios (e.g., \citealt{Bromm01,Simcoe04}). 

\subsection{Ly$\alpha$ forest, the temperature and equation of state of the IGM}\label{subsec:SciGoalLyaForestIGMEoS}

After the cosmic reionization of hydrogen, the remaining small fraction of neutral hydrogen produces a forest of resonant Ly$\alpha$ absorption lines in the background quasar spectra. Individual absorption lines in this Ly$\alpha$ forest could be resolved with high resolution, high S/N, and broad wavelength coverage optical spectra (e.g., \citealt{Schaye00,Becker11,Kim13}), such as those obtained with our Magellan/MIKE and MagE spectra through the HIERACHY program (e.g., Figs.~\ref{fig:ExampleMIKESpec}, \ref{fig:ExampleMagESpec}, \ref{fig:OVI}). One of our major scientific goals is to extract the IGM properties from the highly blended Ly$\alpha$ forest in these high-resolution and high S/N spectra (e.g., \citealt{McQuinn16,Peroux20}).

The broadening of the finest Ly$\alpha$ absorption lines is an effective tracer of the IGM temperature. On large scales the dynamics of the gas, or the broadening of the Ly$\alpha$ line, is dominated by the gravity or the global motion of the system. Conversely, on small scales, the internal pressure of the gas becomes more significant. The gas can be heated by shock or photo-ionization processes. For low density gas in the IGM, where the shock heating is less important, the interplay between photo-ionization heating and adiabatic cooling --- resulting from the expansion of the Universe --- leads to a tight power law relation between the temperature and density of the gas. This relationship is often referred to as the ``equation of state'' of the IGM (e.g., \citealt{Hui97,McDonald01,Gaikwad20}):
\begin{equation}\label{equ:GasEquationOfState}
    T_\delta=T_0(\rho/\bar{\rho})^{\gamma-1}, 
\end{equation}
where $T_\delta$ is the temperature of the gas at a given overdensity $\delta=\rho/\bar{\rho}$, $\bar{\rho}$ is the cosmic mean density, and $T_0$ is the gas temperature at $\bar{\rho}$. Studying the redshift evolution of the parameters of the equation of state ($\gamma$, $T_0$, or $T_\delta$ at different $\rho$) is an important tool for tracking the cosmic reionization history (e.g., \citealt{Ricotti00,Schaye00,McDonald01,Gaikwad20}).

Only with a velocity resolution of $\Delta v\lesssim20\rm~km~s^{-1}$ can we resolve the narrowest Ly$\alpha$ lines that are purely thermally broadened in a typical \ion{H}{1} column density range of $\log N_{\rm H{\scriptscriptstyle~I}}/\rm cm^{-2}\sim12-15$ (e.g., \citealt{Rauch97,Schaye00}). This lower limit on the Ly$\alpha$ line width, described by the Doppler $b$-parameter, caused by thermal broadening, results in a cut-off in the $b$-parameter distribution as a function of \ion{H}{1} density or column density. Following previous studies (e.g., \citealt{Ricotti00,Schaye00,Becker11,Gaikwad20}), we will measure this cut-off at different overdensities (in the form of $b$ or $T_\delta$), as well as the slope $\gamma$ of the IGM equation of state at different redshifts.

Studying the Ly$\alpha$ forest could also be helpful to search for the weakest metal lines using at least two approaches: direct stacking or pixel-by-pixel optical depth analysis (e.g., \citealt{Ellison00}). Since high column density absorbers may be significantly contaminated by the CGM or galactic outflow, measuring metal lines down to the lowest possible column densities is critical for studying the metal enrichment of the IGM. One limitation in detecting these weakest absorbers is the S/N, which could be significantly improved by stacking many sections of the spectra. Although highly contaminated and embedded in the Ly$\alpha$ forest, the Ly$\alpha$ absorber is always the strongest. We could select Ly$\alpha$ absorbers in a certain column density range (e.g., $\log N_{\rm H{\scriptscriptstyle~I}}/\rm cm^{-2}\sim14.0$), where the corresponding metal absorbers (e.g., \ion{C}{4}, with the corresponding $\log N_{\rm C{\scriptscriptstyle~IV}}/\rm cm^{-2}\sim11.0$) might not be directly detectable. By stacking the spectrum at the expected location of the metal absorbers, we may be able to detect them, although potential shifts between the Ly$\alpha$ and metal lines may significantly impact the detection (e.g., \citealt{Ellison00}). 

Another method for probing the weakest metal lines based on the Ly$\alpha$ forest is to analyze the optical depth $\tau$ of each pixel in the Ly$\alpha$ forest along with the corresponding metal absorber pixels (e.g., \citealt{Cowie98,Ellison00}; see also an improved method focusing solely on metal lines in \citealt{Songaila05}). This method could be applied to a broad range of $\tau(Ly\alpha)$, and could also trace down to other Lyman series lines. Compared to the direct stacking method, it has the advantage of being less affected by the redshift offset and the contamination from other absorption features. Using this method, \citet{Ellison00} found an almost constant $\log N({\rm C{\scriptstyle~IV}})/N({\rm H{\scriptstyle~I})}\sim-3$ down to $\tau(Ly\alpha)\sim2-3$, and conclude that weaker absorbers below the detection limit are necessary to reproduce the measured $\tau({\rm C{\scriptstyle~IV}})$.

\subsection{Other Scientific Goals}\label{subsec:SciGoalOthers}

In addition to the major scientific goals focused on studying the IGM/CGM during the \ion{He}{2} EoR using quasar absorption lines, the HIERACHY program also encompasses several additional scientific goals. Some of these goals rely heavily on ongoing or planned follow-up observations that are still far from completion. In this section, we will briefly introduce these goals and present some very limited preliminary results.


\subsubsection{Host galaxies of the \ion{C}{4} absorbers}\label{subsubsec:SciGoalCIVHostGalaxy}

After identifying intervening \ion{C}{4} and other metal or Ly$\alpha$ absorbers in the background quasar spectra, our next step is to search for galaxy candidates associated with these absorbers. Detecting the host galaxies associated with strong \ion{C}{4} absorbers is crucial for understanding the ionization and enrichment sources responsible for the observed \ion{C}{4} absorbers, which in turn sheds light on the cosmic reionization and enrichment history (e.g., \citealt{Diaz14,Diaz15,Cai17b}). Additionally, studying the spatial distribution of the gas traced by the absorbers around the host galax(ies) is also critical in understanding the CGM extension and its interface with the larger-scale IGM/ICM (e.g., \citealt{Werk13}). 

One of the most common methods to search for galaxy candidates within a specific redshift range is to detect LAEs using narrow-band imaging (e.g., \citealt{Shimasaku06,Ouchi08,Ouchi18,TorralbaTorregrosa23}). Most LAEs are young, metal-poor, low-mass star-forming galaxies characterized by weak UV continuum but prominent Ly$\alpha$ emission lines. Although only a small fraction of LAEs host strong AGN (e.g., \citealt{Ouchi08,Ouchi18,Ouchi20,Ono10}), AGN may dominate the LAE population at the high luminosity end (e.g., \citealt{TorralbaTorregrosa23}). LAEs also show weak correlations with other star-forming galaxy tracers (e.g., \citealt{Ito21}), although all these tracers may actually trace the same galaxy overdensities (e.g., \citealt{Harikane19}). The luminosity function (LF) of LAEs can provide insights into the stellar population and initial mass function (IMF) of these early star-forming galaxies, as well as the IGM neutral fraction or the escape fraction of ionizing photons (e.g., \citealt{Ono10,Ouchi18,Konno18,TorralbaTorregrosa23}).

\begin{figure*}[bt]
\centering
\includegraphics[width=1.0\textwidth]{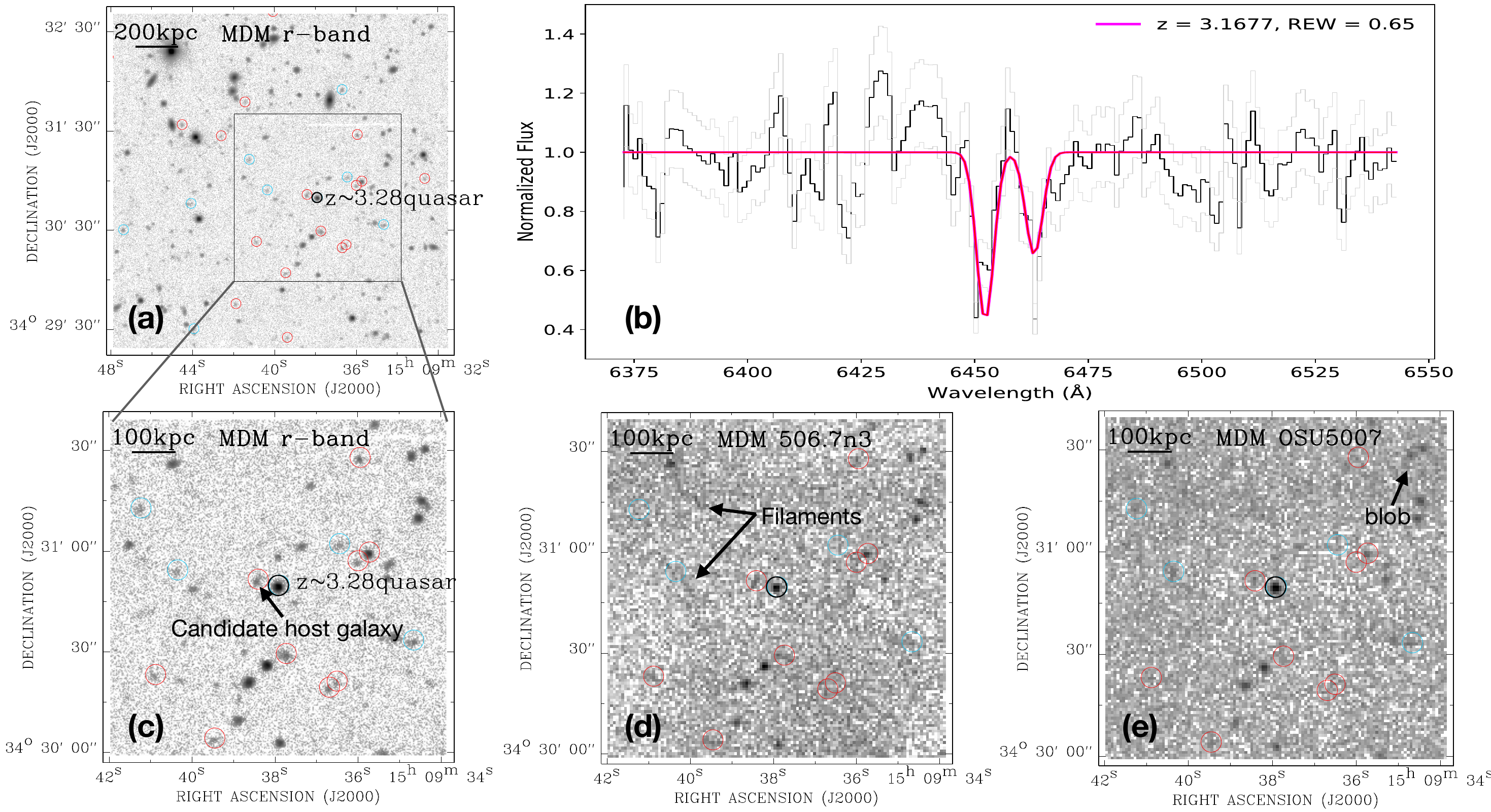}\\
\caption{An example of our imaging observations of the field of the $z\approx3.28$ quasar J150937.9+343049 [panels~(a,c,d,e); the quasar is marked with a black circle). The quasar has a strong \ion{C}{4} absorber with $\log N_{\rm C{\scriptscriptstyle~IV}}/\rm cm^{-2}\approx14.31$ at $z\approx3.17$ [panel~(b); \citealt{Cooksey13}]. Panel~(a) has a FOV of $\sim200^{\prime\prime}\times200^{\prime\prime}$, while the lower three panels are the zoom-in of the $100^{\prime\prime}\times100^{\prime\prime}$ box in (a) centered at the quasar. The two narrow-band images are rebinned to increase the S/N. We also mark a candidate host galaxy of the absorber, as well as some potentially extended features in different panels. The $5~\sigma$ detection limits in the 506.7n3, OSU5007, and $r$-bands are 22.20, 22.84, and 23.83~mag, respectively, all with a total exposure time per filter of 4.5~hours. The red and blue circles are the detected sources brighter in the 506.7n3 ($z\sim3.159-3.183$, around the $z\approx3.17$ \ion{C}{4} absorber) or OSU5007 filters ($z\sim3.106-3.131$), respectively, with a differential magnitude of $\mid\Delta mag({\rm 506.7n3-OSU5007})\mid > 1$ and a broad-band magnitude $<24\rm~mag$. These sources are regarded as candidates of LAEs, but the above selection criteria are arbitrary and not yet quantitatively verified.}\label{fig:J1509MDMimages}
\end{figure*}

At the current stage, we are conducting narrow-band imaging observations primarily with the 1.3m McGraw-Hill and 2.4m Hiltner telescopes at the MDM observatory located at Pitt Peak, Arizona, USA. Using the redshifted [\ion{O}{3}] and a few other narrow-band filters available at the MDM observatory, our focus is on detecting LAEs within a narrow redshift range of $z\sim3.1-3.3$. Since the MDM observatory is located at a latitude of $\rm 31^\circ57^\prime06^{\prime\prime}N$ while the Las Campanas Observatory (LCO), which hosts the Magellan telescopes, is located at $\rm 29^\circ00^\prime57^{\prime\prime}S$, only a small fraction of the HIERACHY quasar sample can be covered by both telescopes. Consequently, in our preliminary test imaging observations, we primarily observe a quasar sample from the northern sky, with strong \ion{C}{4} absorption lines identified in the SDSS spectra \citep{Cooksey13}. Details of the completed observations and the initial results based on them will be presented in follow-up papers. 

We herein present our preliminary test images of the field around the $z\approx3.28$ quasar J150937.9+343049 as an example, which has a strong \ion{C}{4} absorber with $\log N_{\rm C{\scriptscriptstyle~IV}}/\rm cm^{-2}\approx14.31$ at $z\approx3.17$ (Fig.~\ref{fig:J1509MDMimages}). We obtained a broad $r$-band image, as well as two narrow-band images using the 506.7n3 (center wavelength and full width at 50\% peak transmission of $\lambda_{\rm c}=5071~\text{\AA}$, $W_{\rm 50}=29~\text{\AA}$) and OSU5007 ($\lambda_{\rm c}=5010~\text{\AA}$, $W_{\rm 50}=29~\text{\AA}$) filters with the MDM 2.4m telescope. With 4.5~hours of exposure time in each band, we reached a $5~\sigma$ detection limit of 22.20, 22.84, and 23.83~mag in the 506.7n3, OSU5007, and $r$-bands, respectively. This sensitivity is typically sufficient to detect an $L^\star$ star forming galaxy with a stellar mass of $M_*\sim10^{10}\rm~M_\odot$ and star formation age $t_{\rm SF}\lesssim0.5\rm~Gyr$. 

Using the two neighbouring narrow-band filters, we can select LAE candidates based on differential magnitudes between the two bands. The 506.7n3 filter covers Ly$\alpha$ at $z\sim3.159-3.183$, around the $z\approx3.17$ \ion{C}{4} absorber, while OSU5007 covers Ly$\alpha$ at $z\sim3.106-3.131$. A quantitative selection criterion requires a careful calibration of the data, and is still under development. Currently we use an arbitrary value of $\mid\Delta mag({\rm 506.7n3-OSU5007})\mid > 1$ only to demonstrate the feasibility of the method, with sources brighter in either of the two narrow bands highlighted in Fig.~\ref{fig:J1509MDMimages}. Notably, we detected an LAE candidate that is brighter in 506.7n3 projected at $r\approx6.5^{\prime\prime}\approx50\rm~pkpc$ (assuming a redshift of $z=3.17$) from the $z\approx3.28$ background quasar. This LAE candidate is likely the host galaxy of the strong \ion{C}{4} absorber, and follow-up spectroscopic observations are required to confirm this.

In addition to identifying the host galaxies of the strong \ion{C}{4} absorbers, we will also search for any extended Ly$\alpha$ nebulae around individual galaxies using direct narrow-band imaging (see examples of candidate extended features in Fig.~\ref{fig:J1509MDMimages}d,e), or by stacking narrow-band images around identified LAEs. LAEs with extended Ly$\alpha$ envelopes are known as Ly$\alpha$ blobs; some of these structures can be as large as $>100\rm~kpc$ and are referred to as enormous Ly$\alpha$ nebulae (ELAN; e.g., \citealt{Cantalupo14,Borisova16,ArrigoniBattaia19,Ouchi20}). These extended nebulae could represent the direct gas reservoirs associated with the Ly$\alpha$ and metal absorption lines.


\subsubsection{(Proto-)cluster candidates}\label{subsubsec:SciGoalProtoCluster}

The \ion{He}{2} EoR is also a crucial stage for studying the formation and virialization of large-scale gravitationally bound structures, such as (proto-)clusters. There are several methods to search for large-scale structures in the early Universe (see \citealt{Overzier16} and references therein): (1) Search for galaxy overdensities in both spatial and redshift domain via either optical spectroscopy or narrow-band imaging surveys (e.g., \citealt{Ascaso16,Clerc16}). This method has the advantage of relatively high accuracy in spectroscopy or photometric redshift measurements, making it more efficient in identifying gravitationally bound systems such as clusters, some of which may even have X-ray or SZ detections (e.g., \citealt{Clerc16}). However, the detection limit is often high, making this method more suitable for relatively low-redshift objects (e.g., at $z<1$), unless applied to deep but narrow fields (e.g., \citealt{Lilly07,Higuchi19}). (2) Study galaxy clustering in optical broad-band or multi-wavelength wide-field surveys or deep fields (e.g., \citealt{Yang07,Coupon12,Geach17}). This method could typically reach lower detection limits compared to narrow-band or spectroscopy surveys at a similar cost of telescope time. However, because only broad-band data are available, the accuracy of photometric redshifts is insufficient to identify (proto-)clusters. Follow-up multi-object spectroscopy observations are often needed to further identify candidates of large-scale structures (e.g., \citealt{Toshikawa14,Toshikawa16}). In some cases, high angular resolution gravitational lens data can directly identify some cluster candidates (e.g., \citealt{Jaelani20}). (3) Search for LAEs anchored at some high-$z$ objects, such as known bright quasars. The concept is based on the hypothesis that the anchored object tends to reside in massive halos. This method could identify some very high-$z$ proto-clusters (e.g., \citealt{Hu21}), but has two potential disadvantages: First, it is limited to large-scales structures around known objects; Second, since bright quasars are not necessarily associated with proto-clusters that have strong LAE member galaxies (e.g., \citealt{Uchiyama18}), the identification efficiency is not very high. (4) Similar as method (3), but the target large-scale structures are anchored at foreground gaseous absorbers in the spectra of background quasars (e.g., \citealt{Cai17a}). The most commonly adopted absorption line tracer is, of course, the strongest Ly$\alpha$ lines. However, at high-$z$, the strong Ly$\alpha$ forest can make identifying the foreground absorbers difficult. In this project, we propose using \ion{C}{4} absorbers as the primary tracer of large-scale gravitationally bound structures. Compared to Ly$\alpha$, the \ion{C}{4} lines are much easier to identify and measure, and they often trace gas at higher temperatures, making them more likely to be associated with gravitationally bound systems. 

As introduced in \S\ref{subsubsec:SciGoalCIVHostGalaxy}, we are using narrow-band images covering the Ly$\alpha$ emission line and neighbouring bands to identify LAEs. This method can also be used to identify galaxy overdensities, making it useful for finding candidates of (proto-)clusters or other large-scale structures. We will particularly focus on strong and/or complex \ion{C}{4} absorbers, which can often be decomposed into multiple components spread across a velocity range of $\Delta v\sim100-400\rm~km~s^{-1}$. These absorbers typically have high column densities of $\log N{\rm(C{\small~IV})/cm^2}\gtrsim14$, suggesting they may be produced by the intragroup medium (IGrM) of galaxy groups or the ICM of galaxy (proto-)clusters. Limited by the available filters at the MDM observatory, most of our current narrow-band imaging observations cover only the LAEs at $z\sim3.2$ (e.g., Fig.~\ref{fig:J1509MDMimages}). This redshift represents the earliest stage where some large-scale structures become gravitationally bound or ``mature'' enough to contain a significant amount of hot gas detectable in X-ray or SZ observations (e.g., \citealt{WangT16,Tozzi22}). 

Using LAEs as tracers of large-scale structures could be biased toward low-mass star-forming galaxies, which may not be the dominant galaxy populations responsible for the \ion{C}{4} (or other gas phases) reservoirs (e.g., \citealt{Ouchi08,Ouchi18,Ouchi20,Ono10}). Therefore, we need multi-wavelength multi-object spectroscopy observations to further identify and study many other galaxies with different emission lines in the selected LAE overdensity fields. These multi-wavelength observations could include multi-object spectroscopy in the optical band (e.g., \citealt{Diaz15,Lemaux18,Harikane19}) or radio interferometry observations covering different lines (e.g., redshifted CO~$J=1-0$ in cm wave band, [\ion{C}{2}]~$\lambda 157\rm~\mu m$ in mm wave band, etc.; \citealt{WangT16,Wang24b}). Additionally, these multi-object spectroscopy observations are essential for measuring the velocity dispersion of the system, which is crucial for identifying gravitationally bound large-scale structures, such as (proto-)clusters (e.g., \citealt{Toshikawa14,Toshikawa16,WangT16,Cai17a,Hu21,Shi21}). We will prioritize our follow-up multi-object spectroscopy observations on pre-selected systems with the most complex \ion{C}{4} absorption structures and identified LAE overdensities. Following the quasar absorption line and narrow-band imaging observations, these multi-wavelength observations have just begun as part of the HIERACHY program. As of the submission of this paper, we have obtained new \emph{JVLA} observations (in the 2024A semester) of three of our quasars fields visible from the northern sky to detect the redshifted CO~$J=1-0$ emission from star forming galaxies in the field.

The ultimate confirmation of virialized galaxy clusters is the detection of their hot ICM, either through X-ray or SZ signal observations (e.g., \citealt{WangT16,Tozzi22}). X-ray emission from hot gas is proportional to its density and metallicity, but is less sensitive to its temperature. In contrast, the strongest thermal SZ signal is proportional to the density and temperature of the hot gas, but is less affected by its metallicity. Therefore, joint X-ray and SZ observations of the hot ICM are critical for accurately estimating its physical and chemical properties. Furthermore, X-ray observations with \emph{Chandra} and/or \emph{XMM-Newton} often have good angular resolution, allowing for the separation of AGN from diffuse gas. However, the sensitivity of these observations declines rapidly with increasing redshift. On the other hand, SZ observations, although not significantly affected by the redshift, are often highly contaminated by bright point sources in the field. Combining these two methods, most existing observations are only able to detect the hot ICM from massive clusters at $z\lesssim2$ (e.g., \citealt{Carlstrom02,Bleem15,Bartalucci17,Bartalucci19}). Either extra-deep observations or new telescopes are needed to detect the hot ICM from the first virialized clusters at our desired redshift of $z\gtrsim3$. Our pre-selection of high velocity dispersion (proto-)cluster candidates will provide a reliable candidate sample for searching for these earliest massive gravitationally bound systems.


\subsubsection{Outflows and inflows associated with quasars}\label{subsubsec:SciGoalOutflowInflow}

The high-resolution rest-frame UV spectra obtained in the HIERACHY program also provide a powerful tool for studying AGN outflows and inflows during the \ion{He}{2} EoR. AGN outflows can be studied through either emission lines or absorption lines. Many high-$z$ quasars have broad blueshifted emission lines from high and intermediate ionization ions, indicating that fast outflows are quite common in these brightest objects at the early ages of the universe (e.g., \citealt{Yu21}). The corresponding AGN feedback via these outflows could play an important role in the cosmic reionization and the co-evolution of the galactic ecosystems (e.g., \citealt{Heckman14}). In addition to the broad blueshifted emission lines, AGN outflow could also be detected in absorption lines (Fig.~\ref{fig:OVI}). Resolving and measuring these often complex multi-component absorption systems, in many cases also from multi-ions, can help determine many physical parameters of the AGN outflow (e.g., \citealt{Chen18,Wang18,Wang21,Xu18,Byun22,Yang23}). To minimize contamination from AGN outflows in our absorption line study of the intervening IGM, we have excluded \ion{C}{4} absorbers within $\Delta v \leq 5000\rm~km~s^{-1}$ blueward of the expected location of the \ion{C}{4} emission line of the quasar (\S\ref{subsec:MagellanMIKE}). Absorbers in this wavelength range will not be included in our studies of the IGM \ion{C}{4} absorbers (e.g., HIERACHY~III), but they remain valuable for identifying and studying AGN outflows, some of which will be published in companion papers (e.g., HIERACHY~I \& IV).

In addition to the outflows traced by blueshifted emission and absorption lines, some AGN also exhibit redshifted absorption lines identified on the wings of broad emission lines (e.g., \citealt{Hall13,Chen22}). These redshifted absorption lines trace AGN inflows, which are possibly metal-enriched dense clouds falling back either from the CGM or the AGN outflow (e.g., \citealt{Gaspari13}). Measuring the physical and chemical properties of these AGN inflows will help us better understand the accretion processes of SMBHs (e.g., \citealt{Hu08,Gaskell13,Gaskell16,Grier17}. Unlike the absorption lines from outflows, which can be either broad or narrow, most inflows exhibit relatively narrow absorption components. However, there are cases where broad absorption lines (BALs) ($\Delta v > 2000\rm~km~s^{-1}$) or mini-BALs ($500 \rm~km~s^{-1}< \Delta v < 2000\rm~km~s^{-1}$) are redshifted relative to the quasar (e.g., \citealt{Hall13,Shi17,Zhang17,Zhou19}). A key challenge in identifying quasar inflows is the uncertainty in determining the redshift of the quasar. For example, the blueshifted \ion{C}{4} absorption line shown in Fig.~\ref{fig:OVI} appears to fall on the red wing of the quasar's \ion{C}{4} emission line. 

At the redshift of the \ion{He}{2} EoR, our optical spectra taken with Magellan/MIKE or MagE only cover the rest-frame far-UV band, which primarily consists of strong quasar emission lines from high ions (e.g., \ion{N}{5}, \ion{Si}{4}, \ion{C}{4}). These lines are often heavily contaminated by strong outflows (e.g., \citealt{Yu21}), making it difficult to determine the quasar redshift based on them. Only a subset of our sample quasars have near-IR spectra covering the rest-frame \ion{Mg}{2} or Balmer lines, which are generally less affected by quasar outflows. We are also conducting near-IR spectroscopy observations to better determine the redshifts of these quasars. These observations are also crucial for obtaining unbiased SMBH mass estimate using the single-epoch method (e.g., \citealt{Shen12}).

For some bright quasars with strong outflows or inflows, we are also conducting multi-epoch medium- to high-resolution spectroscopy observations to study the variations in the velocity and strength of the absorption lines (e.g., \citealt{Yi19a,Yi19b}). Although these follow-up observations are not directly related to the major scientific goals of the HIERACHY program, they are important for understanding the mechanisms behind the launching and acceleration/deceleration of quasar outflows and inflows.

\section{Summary and Prospect} \label{sec:Summary}

In this paper, we introduce the scientific motivation, observational design, major scientific goals, and some examples of initial data products from the HIERACHY program. As of the submission of this paper, we have completed Magellan/MIKE high-resolution spectroscopy observations of 26 quasars at $z\approx3.9-5.2$, most of which achieve a resolution of $R\approx32,000$ with a signal-to-noise ratio of $\rm S/N_{25\%}\gtrsim20\rm~pixel^{-1}$. This typically allows us to reach a \ion{C}{4} column density detection limit of $\log N{\rm(C{\small~IV})/cm^2}\lesssim12.7$ for most pixels in the redshift range of $z\sim3-5$. Many pixels with a better $\rm S/N$ enable us to probe the IGM down to as low as $\log N{\rm(C{\small~IV})/cm^2}\sim12$ in the expected \ion{He}{2} EoR, given that C$^{3+}$ and He$^{+}$ have comparable ionization potentials. We have also completed medium-resolution spectroscopy observations of 29 quasars with Magellan/MagE, achieving a typical resolution of $R\approx7,000$ with a signal-to-noise ratio of $\rm S/N_{25\%}\gtrsim10\rm~pixel^{-1}$. These MagE observations will significantly enhance the statistics at the high-$N_{\rm C{\small~IV}}$ end. We present some initial products from our spectroscopy and imaging observations, while detailed analyses of the different datasets and the major scientific findings will be presented in a series of follow-up papers.

High-resolution spectroscopy observations of quasars at higher redshifts or with a higher resolution and/or S/N are of particular scientific interest for studying the weakest intervening IGM/CGM absorbers over a broad redshift range (e.g., \citealt{Yang20b,Jin23}). These observations are typically very time-intensive with current 6-10~m ground-based optical telescopes (e.g., \citealt{Ellison00}). In recent years, deeper quasar surveys at high-$z$ (e.g., \citealt{WangF16,Yang16,Cristiani23,Fan23}) have identified more background sources available to probe the cosmic reionization history. These objects are typically fainter than those studied in our HIERACHY program, requiring larger telescopes to obtain high-quality spectra. In approximately 10 years, the next generation of $\sim30\rm~m$ telescopes will begin operations. For instance, the ELT/ANDES could typically reach a $\rm S/N\sim30~pixel^{-1}$ at a resolution of $R\sim100,000$ for a $z=7$ quasar with a J-band magnitude $mag_{\rm J}\gtrsim20$ \citep{Marconi22,DOdorico23}. Such high-quality spectra will be critical for probing cosmic reionization processes in much greater detail, extending our understanding back into the hydrogen EoR. 

New observations are also needed to probe the early formation of large-scale gravitationally bound systems, i.e., (proto-)clusters. The most efficient way to study (proto-)cluster member galaxies will involve optical IFU or multi-object spectroscopy, or radio interferometry fine-tuned to the frequency of specific emission lines. In the future, the next generation of $\sim30\rm~m$ telescopes will enable us to obtain spectra of much fainter cluster member galaxies or background sources. For example, the ELT/MOSAIC could typically achieve a $\rm S/N\gtrsim5~pixel^{-1}$ at a resolution of $R\sim5,000$ for a $z\sim3-4$ galaxy with a rest frame UV magnitude of $\sim25.5\rm~mag$ \citep{Puech18,Japelj19}. This capability is critical not only for measuring the emission line properties of cluster member galaxies and the velocity dispersion of the (proto-)cluster, but also for conducting absorption line studies of foreground large-scale gaseous structures using high-density background sources, which are likely to be LBGs rather than quasars. The combination of emission line studies (with the multi-IFU mode mIFU of MOSAIC) and absorption line studies will be invaluable for investigating gaseous structures on different physical scales. In addition to studying cluster member galaxies and the cool CGM/IGrM/ICM, new X-ray and SZ observations are also anticipated to detect the earliest signals from the hot ICM. Most the existing X-ray and SZ observations only detect mature clusters with hot ICM at $z\lesssim2$ (e.g., \citealt{Carlstrom02,Bleem15,Bartalucci17,Bartalucci19}). The highest-$z$ galaxy cluster with confirmed virialized hot ICM is at $z\approx2.5$ \citep{WangT16}. We expect to identify some (proto-)cluster candidates with the highest velocity dispersions at $z\gtrsim3$ through the HIERACHY program. These objects will be prime candidates for follow-up X-ray and SZ observations to search for the first virialized ICM. Existing telescopes, such as \emph{Chandra} and \emph{XMM-Newton} in X-ray, and \emph{IRAM 30m}, \emph{LMT}, and \emph{GBT} in radio, are already sensitive enough to detect the most massive galaxy clusters at such high redshifts. Future telescopes, such as \emph{Athena} (e.g., \citealt{Nandra13}) and \emph{AXIS} in X-ray, and \emph{CMB-S4} in radio (e.g., \citealt{Carlstrom19}), will be even more powerful in conducting spatially resolved studies of the hot ICM and in detecting the ICM of lower-mass galaxy clusters at even higher redshifts.

\section*{Acknowledgements}

The authors would like to acknowledge Prof. Ian U. Roederer from the North Carolina State University, Dr. Fengwu Sun from the University of Arizona for their contributions in observations, data analysis, and/or scientific discussions. J.T.L. acknowledges the financial support from the National Science Foundation of China (NSFC) through the grants 12273111 and 12321003, and also the science research grants from the China Manned Space Project. T.F. is supported by the National Key R\&D Program of China under No. 2017YFA0402600, the National Natural Science Foundation of China under Nos. 11890692, 12133008, 12221003, and the science research grant from the China Manned Space Project with No. CMS-CSST-2021-A04. X. W. is supported by the Fundamental Research Funds for the Central Universities, the CAS Project for Young Scientists in Basic Research Grant No. YSBR-062, and the Xiaomi Young Scholars Fellowship. Y.Y.S. acknowledges support from the Dunlap Institute that is funded through an endowment established by the David Dunlap family and the University of Toronto.

\bibliographystyle{mnras}
\bibliography{HIERACHY_II}

\end{document}